\documentclass[nojss
]{jss}

\usepackage[utf8]{inputenc}

\author{
Victor Freguglia\\University of Campinas \And Nancy Lopes
Garcia\\University of Campinas
}
\title{Inference tools for Markov Random Fields on lattices: The
\proglang{R} package \pkg{mrf2d}}

\Plainauthor{Victor Freguglia, Nancy Lopes Garcia}
\Plaintitle{Inference tools for Markov Random Fields on lattices: The R
package mrf2d}
\Shorttitle{\pkg{mrf2d}: Markov Random Fields on lattices}

\Abstract{
Markov random fields on two-dimensional lattices are behind many image
analysis methodologies. \pkg{mrf2d} provides tools for statistical
inference on a class of discrete stationary Markov random field models
with pairwise interaction, which includes many of the popular models
such as the Potts model and texture image models. The package introduces
representations of dependence structures and parameters, visualization
functions and efficient (\proglang{C++} based) implementations of
sampling algorithms, common estimation methods and other key features of
the model, providing a useful framework to implement algorithms and
working with the model in general. This paper presents a description and
details of the package, as well as some reproducible examples of usage.
}

\Plainkeywords{Markov random fields, image analysis, R, Gibbs random
fields, Potts model, texture}

\Address{
    Victor Freguglia\\
  University of Campinas\\
  Institute of Mathematics, Statistics and Scientific Computation\\
  E-mail: \email{victorfreguglia@gmail.com}\\
  URL: \url{https://freguglia.github.io}\\~\\
      Nancy Lopes Garcia\\
  University of Campinas\\
  Institute of Mathematics, Statistics and Scientific Computation\\
  E-mail: \email{nancyg@unicamp.br}\\
  
  }

\usepackage{amsmath} \usepackage{amssymb} \usepackage{booktabs} \usepackage{longtable} \usepackage{float} \usepackage{bbm} \usepackage[ruled,vlined]{algorithm2e}   

\begin{document}

\hypertarget{introduction}{%
\section{Introduction}\label{introduction}}

A Markov Random Field (MRF) is a generalization of the well-known
concept of a Markov Chain where variables are indexed by vertices of a
graph instead of a sequence and the notion of memory is substituted by
the neighborhood (edges) of that graph. Markov Random Fields on
lattices, or more generally, Gibbs distributions, have been studied in
Statistical Mechanics as models for interacting particle systems. They
range from the very simple Ising model (or its generalization Potts
model) with pairwise nearest-neighbor interaction to models with more
complex interaction types, presenting long-range and/or higher-order
interaction. For an introduction to the subject we refer to
\citet{liggett2012interacting} and references therein.

A finite 2-dimensional lattice is a direct representation of pixel
positions on a digital image. \citet{geman1984stochastic} make an
analogy between image models and statistical mechanics systems,
introducing probability-based computational methods for image
restoration under a specific type of noise. Higher-order dependence
structures are also described, for example, interactions with pixels
more distant than nearest-neighbors. \citet{cross1983markov} use MRFs
with special interaction structures to model texture images.

Many modern image analysis methodologies in statistics and machine
learning are grounded on Markov Random Field theory and the local
dependence characteristic of image data. Common tasks in image analysis
involve image segmentation
\citep{zhang2001segmentation, kato2006markov, roche2011convergence, cao2018hyperspectral, ghamisi2018new},
texture synthesis
\citep{gimel1996texture, freeman2011markov, versteegen2016texture} and
statistical modeling
\citep{derin1987modeling, guillot2015statistical, freguglia2019hidden}
all of which can be achieved with the use of MRFs. Some basic references
are \citet{blake2011markov} and \citet{kato2012markov}.

In this paper, when we refer to a MRF, we consider the particular case
where variables are indexed by points of a 2-dimensional lattice, not a
general graph structure. The regular grid naturally creates a spatial
structure and notions of distance and direction for the variables,
allowing models to be specified based on this spatial structure
\citep[see][ for examples]{besag1974spatial}.

Parametric inference based on maximum likelihood for such models is
difficult, even for the simple models, because of the intractable
constant that appears in the likelihood. Inference for the simplest
non-trivial case of the Ising model was first studied by
\citet{pickard1987inference} and continues to present challenges, see
for example \citet{bhattacharya2018inference}. On the other hand, while
there is a continuous development of methodologies used in MRFs in the
theoretical field, implementing new algorithms is a challenge in
practice, mostly due to the high-dimensionality of the problem and the
complexity of the data structures required to represent the data in this
type of problem. An overview of this topic, mainly from the Bayesian
perspective, can be found in \citet{winkler2012image}.

Most methodologies developed are based on Monte-Carlo Markov Chain
methods, thus simple tasks like evaluating pairs of pixels or sampling
individual pixels need to be repeated millions or billions of times in
iterative methods, depending on the image size, making an efficient
implementation of such methods one of the main demands for researchers
of the topic.

\proglang{R} \citep{ref_rlang} is one of the most used programming
languages among Statistics researchers, what makes the existence of good
packages important for any field of Statistics. In a general context
(considering the definition of MRFs with general undirected graphs),
packages like \pkg{graph} \citep{pkggraph} and \pkg{network}
\citep{pkgnetwork} provides tools for representation and manipulation of
graph structures which can be used for constructing and visualizing
graph-based models. Different versions of graph-based MRFs appear in
many packages. For example, the \pkg{CRF} package \citep{CRF} has
inferential tools Markov random fields with pairwise and unary
interactions and their hidden MRF version, \pkg{MRFcov}
\citep{clark2018unravelling} allows inference for the interaction
parameter of between nodes of a graph considering covariates and
\pkg{gamlss.spatial} \citep{de2018gaussian} allows fitting Gaussian
Markov random fields in a spatial context, similar to \pkg{INLA}
\citep{INLA} and \pkg{mgcv} \citep{mgcv}.

Outside of the \proglang{R} ecosystem, there are powerful software in
\proglang{C++} used for image analysis related to Markov random fields,
such as the \pkg{DGM} library \citep{DGM} and \pkg{densecrf}
\citep{krahenbuhl2011efficient}, which also has a \proglang{Python}
wrapper \citep{pydensecrf}, and can be used for a variety of tasks and
use extremely efficient computational methods.

For discrete MRFs on lattice data, closer to what is proposed in
\pkg{mrf2d}, there are some \proglang{R} packages available. The
\pkg{potts} package \citep{ref_pottspkg}, implements simulation
algorithms and parameter estimation via Composite-Likelihood for a Potts
model with nearest-neighbor interactions only. \pkg{PottsUtils}
\citep{ref_pottsutils} also implements simulation and tools for
computing normalization constants in one, two and three-dimensional
Potts model. The package \pkg{bayesImageS} \citep{bayesimages} provides
Bayesian image segmentation algorithms considering Gaussian mixtures
driven by Hidden Potts models with slightly more complex interaction
neighborhood. \pkg{GiRaF} \citep{stoehr2016giraf} allows calculation on,
and sampling from general homogeneous Potts model. The \pkg{Pottslab}
\citep{pottslab} package for \proglang{MATLAB} also provides image
segmentation algorithms using the Potts model, including for
multivariate-valued data.

Although the available packages for discrete-valued MRFs offer efficient
implementations of their methods, they do not provide an interface that
allows simple extensions to different cases, for example, different
interaction types for different positions and sparse long-range
interaction neighborhoods. Some of the algorithms used also rely on
specific characteristics of the specific setups they consider and cannot
be applied more generally.

The \pkg{mrf2d} package \citep{mrf2d_cran} provides a complete framework
for statistical inference on discrete-valued MRF models on 2-dimensional
lattice data, where all the elements used by algorithms (such as
conditional probabilities, pseudo-likelihood function, simulation,
sufficient statistics and more) are available for the user, as well as
many built-in model fitting functions.

The package uses the model described in \citet{freguglia2019hidden} as a
reference. Many other models, such as the Potts model and auto-models,
are particular cases of our model obtained by including restrictions to
the parameters or using specific interacting neighborhoods. These
neighborhoods can be freely specified within the package and 5 families
of parameter restrictions are available to cover the particular cases.

\pkg{mrf2d} is available on CRAN and as a development version in its git
repository. These versions can be installed with

\begin{CodeChunk}

\begin{CodeInput}
R> install.packages("mrf2d") # CRAN version
R> devtools::install_github("Freguglia/mrf2d") # Development version
R> library("mrf2d")
\end{CodeInput}
\end{CodeChunk}

This paper is organized as follows. Section \ref{model-description}
describes the model considered in \pkg{mrf2d}, Section
\ref{using-the-package} presents the main functionalities of the package
and details of the implementation, which are illustrated by examples in
Section \ref{sec:examples}. We finish with a discussion in Section
\ref{discussion}.

\hypertarget{model-description}{%
\section{Model description}\label{model-description}}

Let
\(\mathcal{L} \subset \{{\mathbf i}= (i_1, i_2) \in \mathbb{N}^2 \}\) a
finite set of locations in a two-dimensional lattice region and
\(\mathbf{Z} = \{Z_{\mathbf i}\}_{{\mathbf i}\in \mathcal{L}}\) a field
of random variables indexed by those locations.

The main purpose of \pkg{mrf2d} is to provide a general framework for
Markov random field models which satisfy the following assumptions:

\begin{description}
\item[(a) Finite support] Each $Z_{\mathbf i}$ can take values in 
$\mathcal{Z} = \{0, \dots, C \}$ for some finite $C > 0$.
\item[(b) Pairwise interactions] The probability of a complete
configuration $\mathbb{P}(\mathbf{Z} = \mathbf{z})$ can be decomposed
into a product of functions of the pairs 
$(z_{\mathbf i}, z_{\mathbf j}), {\mathbf i}\neq {\mathbf j}\in \mathcal{L}$.
\item[(c) Homogeneous interactions] The interaction between to pixels
${\mathbf i}$ and ${\mathbf j}$ is the same as for the pixels ${\mathbf i}'$ and ${\mathbf j}'$ if
${\mathbf i}- {\mathbf i}' = {\mathbf j}- {\mathbf j}'$, i.e., the interactions depend on the relative
position of a pair of pixel, not on their position in the lattice.
\end{description}

These assumptions are satisfied by most commonly used models in image
processing.

We use the representation in \citet{freguglia2019hidden} which expresses
the probability distribution of the random field in the form of the
exponential family and introduce additional constraints to parameter
space and/or different dependence structures to include particular
features of the model under study.

\hypertarget{homogeneous-markov-random-field-with-pairwise-interactions}{%
\subsection{Homogeneous Markov Random Field with pairwise
interactions}\label{homogeneous-markov-random-field-with-pairwise-interactions}}

MRF models are characterized by their conditional independence property.
Let \(\mathcal{N}\) a neighborhood system on \(\mathcal{L}\), then
\(\mathbf{Z}\) is a \emph{Markov random field} with respect to
\(\mathcal{N}\) if \(Z_{\mathbf i}\) given its neighbors
\(\mathbf{Z}_{\mathcal{N}_{\mathbf i}}\) is conditionally independent
from all other variables \begin{equation}\label{markov_prop}
\mathbb{P}(Z_{\mathbf i}= z_{\mathbf i}| \mathbf{Z}_{-{\mathbf i}}) = 
\mathbb{P}(Z_{\mathbf i}= z_{\mathbf i}| \mathbf{Z}_{\mathcal{N}_{\mathbf i}}), \hspace{1cm} 
{\mathbf i}\in \mathcal{L},
\end{equation} where \(\mathbf{Z}_{-{\mathbf i}}\) denotes the set of
variables \(\{Z_{\mathbf j}, {\mathbf j}\neq {\mathbf i}\}\).

To start defining MRFs in an image processing context, a location of the
lattice \({\mathbf i}\in \mathcal{L}\) will be referred as a \emph{pixel
\({\mathbf i}\)} and an observed value of the variable
\(z_{\mathbf i}\in \mathcal{Z}\) as \emph{pixel value} or \emph{color}.

We denote \(\mathcal{R} \subset \mathbb{Z}^2\) a set of interacting
relative positions such that, for no pair of elements
\({\mathbf r}, {\mathbf r}' \in \mathcal{R}\) we have
\({\mathbf r}' = -{\mathbf r}\) (no position in \(\mathcal{R}\) is a
reflection of another). Based on \(\mathcal{R}\), we can construct a
neighborhood system (interaction structure) \(\mathcal{N}\) in such way
that the set of neighbors of site \({\mathbf i}\), \(\mathcal{N}_i\) can
be represented by a graph with vertices \(\mathcal{L}\) where there is
an edge connecting \({\mathbf i}\) and \({\mathbf j}\) if, and only if,
\({\mathbf j}= {\mathbf i}\pm {\mathbf r}\). For example, a
nearest-neighbor structure corresponds to
\(\mathcal{R} = \{ (1,0), (0,1) \}\).

Given an interaction structure \(\mathcal{R}\), for any relative
position \({\mathbf r}\in \mathcal{R}\) the interactions associated to
that relative position are characterized by a map
\(\theta_{\mathbf r}(\cdot,\cdot)\),
\(\theta_{\mathbf r}: \mathcal{Z}^2 \rightarrow \mathbb{R}\). For
\(a, b \in \mathcal{Z}\), the value \(\theta_{\mathbf r}(a,b)\) is
called a potential.

The model in \pkg{mrf2d} considers a neighborhood system \(\mathcal{N}\)
that connects pairs of pixel positions \(i,j\) such that
\(i - j \in \mathcal{R}\). Under assumptions (a), (b) and (c), the
Hammersley-Clifford theorem \citep{hammersley1971markov} implies that
the probability function for \(\mathbf{Z}\) belongs to the exponential
family and can be described by a set of natural parameters
\(\boldsymbol{\theta} = \{\theta_{\mathbf r}(a,b), {\mathbf r}\in \mathcal{R}, a,b, \in \mathcal{Z} \}\),
\begin{equation}
  \label{prob_expression}
  \mathbb{P}(\mathbf{Z} = \mathbf{z}) = 
  \frac{1}{\zeta_{\theta}} e^{H(\mathbf{z}, \boldsymbol{\theta})},
\end{equation} where \begin{equation}
  \label{eq_hamiltonian}
  H(\mathbf{z}, \boldsymbol{\theta}) = 
  \sum_{r \in \mathcal{R}} \sum_{i,j \in \mathcal{L}}
    \theta_{r}(z_i, z_j) \mathbbm{1}_{(j = i+r)} 
  \text{ and }
  \zeta_\theta = \sum_{\mathbf{z}'} e^{H(\mathbf{z}',\boldsymbol{\theta})}.
\end{equation}

\begin{figure}[ht]
  \centering
  \includegraphics[width=1\linewidth]{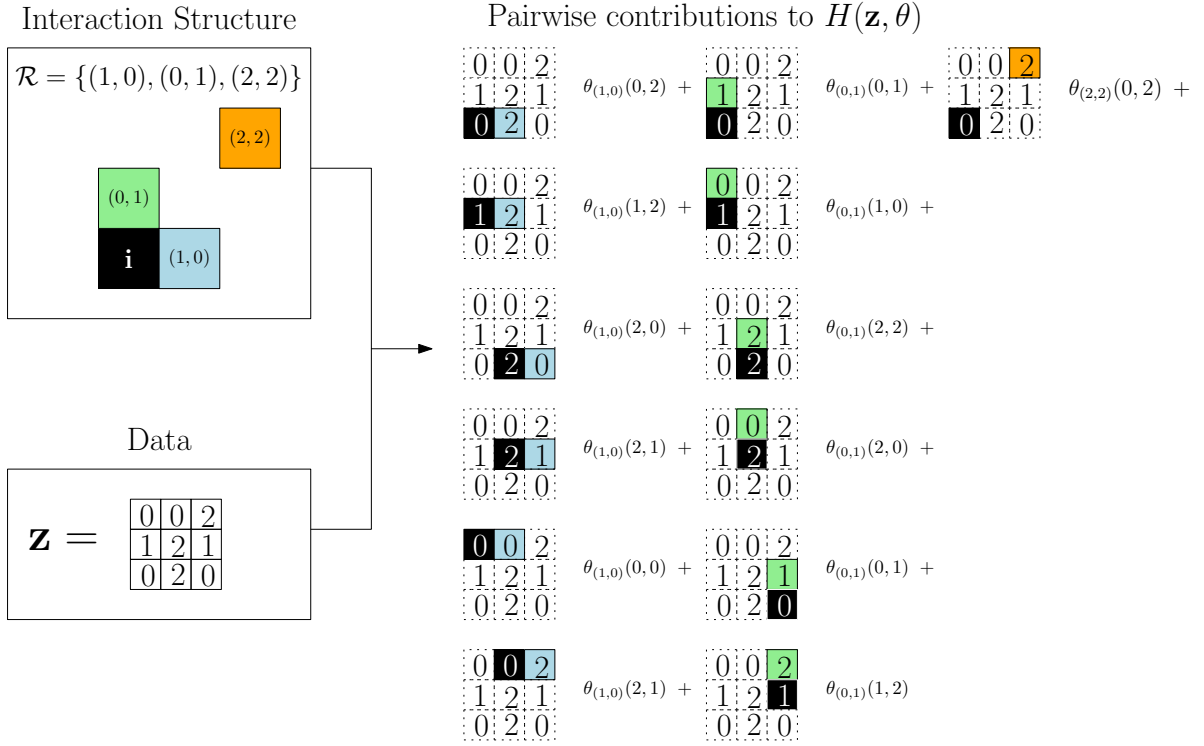}
  \caption{Example of interaction structure with three relative positions and 
example field on a 3 by 3 lattice (left) and contributions of each interacting 
pair to $H(\mathbf{z}, \boldsymbol\theta)$(right).}
  \label{fig_hamilt}
\end{figure}

Figure \ref{fig_hamilt} illustrates how the function
\(H(\mathbf{z},\boldsymbol{\theta})\) is computed for an example
interaction structure \(\mathcal{R}\) and a field \(\mathbf{z}\).

Note that adding a constant a constant \(c_{\mathbf r}\) to the
potentials associated with a relative position
\({\mathbf r}\in \mathcal{R}\) results in the same probability because
the constant cancels when dividing by \(\zeta_{\boldsymbol{\theta}}\).
Thus, constraints for the potentials \(\theta_{\mathbf r}(a,b)\) are
necessary to obtain identifiability in the model. We consider
\(\theta_{\mathbf r}(0,0) = 0\) for all relative positions
\({\mathbf r}\), which ensures identifiability and also gives an
interpretation for interactions in terms of the pair \((0,0)\):
\(\theta_{\mathbf r}(a,b) < 0\) (resp. \(>0\)) means that the pair
\((a,b)\) is \textbf{less} (resp. \textbf{more}) likely to appear in a
pair with relative position \({\mathbf r}\) than \((0,0)\).

\hypertarget{potts-model-as-a-particular-case}{%
\paragraph{Potts Model as a particular
case}\label{potts-model-as-a-particular-case}}

The Potts model \citep{potts1952} is one of the most important MRF model
used in image segmentation because it can assign higher probability for
equal-valued pairs of nearest-neighbors, creating large regions of
pixels with the same values. The model has a single parameter \(\phi\)
that is interpreted as the inverse temperature in a mechanical
statistics context.

A standard Potts Model can be expressed as (\ref{prob_expression}) with
the function \(H(\mathbf{z},\boldsymbol{\theta})\) taking the form
\begin{equation}\label{H_potts}
\phi \sum_{({\mathbf i}, {\mathbf j}):||{\mathbf i}- {\mathbf j}|| = 1} \mathbbm{1}_{(z_{\mathbf i}\neq z_{\mathbf j})}.
\end{equation}

Assumptions (a), (b) and (c) are satisfied, thus, we can rewrite
(\ref{H_potts}) in terms of an interaction structure \(\mathcal{R}\) and
potentials \(\boldsymbol{\theta}\) by noticing

\begin{itemize}
\item The set ${\mathbf i}, {\mathbf j}: ||{\mathbf i}- {\mathbf j}|| = 1$ are vertical and horizontal pairs
of neighbors, therefore, the interaction structure $\mathcal{R}$ is the
set $\{(1,0), (0,1)\}$.
\item The potential $\theta_{\mathbf r}(a,b)$ is equal to $\phi$ if $a \neq b$ and
$0$ otherwise, regardless of ${\mathbf r}$. The constraint $\theta_{\mathbf r}(0,0) = 0$
is satisfied in this definition. Therefore, we have
the parameter restriction
$$\theta_{\mathbf r}(a,b) = \phi \mathbbm{1}_{(a \neq b)}$$ 
for all ${\mathbf r}\in \mathcal{R}$.
\end{itemize}

This parameter restriction corresponds to the \code{"onepar"} family
described in Section \ref{sec:apndx_par}.

\hypertarget{important-elements-of-the-model}{%
\subsection{Important elements of the
model}\label{important-elements-of-the-model}}

The main inference challenge for MRFs lies in the normalizing constant
\(\zeta_{\boldsymbol\theta}\) appearing in (\ref{prob_expression}). It
cannot be evaluated in practice as it requires summing over
\(\mathcal{Z}^{|\mathcal{L}|}\) possible field configurations and there
is no analytical expression for it, except for trivial cases, leading to
an intractable likelihood.

Being unable to evaluate the likelihood function hinders the use of most
statistical methods. Inference under intractable likelihoods have been
developed over the years. The main studies involve using conditional
probability-based functions, like pseudo-likelihood \citep[for
example]{jensen1994} and Monte-Carlo methods \citep[for
example]{geyer1992, moller2006}.

Although there is a wide variety of inferential methods available, most
of them are built using the same pieces of the model. Thus, having
access to each of these pieces is necessary to implement algorithms. We
highlight important characteristics of the model available in
\pkg{mrf2d} that are used by inference methods.

\hypertarget{conditional-probabilities}{%
\paragraph{Conditional Probabilities}\label{conditional-probabilities}}

A consequence of the Markov property (conditional independence) is a
simple expression for conditional probabilities.
\(H(\mathbf{z}, \boldsymbol\theta)\) is a sum of terms that only depends
on pairs of pixel values, which implies that all terms not involving
position \({\mathbf i}\) cancel out when evaluating
\(\mathbb{P}(Z_{\mathbf i}|\mathbf{Z}_{-{\mathbf i}})\). Define the part
of the sum that involves the pixel in position \({\mathbf i}\) as
\begin{equation}
  h_{\mathbf i}(k|\mathbf{z}) = 
  \sum_{{\mathbf r}\in \mathcal{R}}^{({\mathbf i}+{\mathbf r}) \in \mathcal{L}}
    \theta_{\mathbf r}(k, z_{({\mathbf i}+{\mathbf r})}) +
    \sum_{{\mathbf r}\in \mathcal{R}}^{({\mathbf i}-{\mathbf r}) \in \mathcal{L}}
    \theta_{\mathbf r}(z_{({\mathbf i}-{\mathbf r})},k).
\end{equation}

The conditional probability of \(Z_{\mathbf i}= k\) given all other
locations,
\(\mathbb{P}(Z_{\mathbf i}= k|\mathbf{Z}_{\mathcal{N}_{\mathbf i}})\),
is then given by the standard softmax of
\(h_{\mathbf i}(k|\mathbf{z})\), \begin{equation}
  \label{eq_condprob}
  \mathbb{P}(Z_{\mathbf i}= k|\mathbf{Z}_{-{\mathbf i}} = \mathbf{z}_{\mathcal{N}_{\mathbf i}}) =
  \frac{
    e^{h_{\mathbf i}(k|\mathbf{z})}
  }{
    \sum_{k'} e^{h_{\mathbf i}(k'|\mathbf{z})}
  }.
\end{equation}

\hypertarget{pseudo-likelihood-function}{%
\paragraph{Pseudo-likelihood
function}\label{pseudo-likelihood-function}}

The pseudo-likelihood function
\citep{besag1974spatial, besag1975statistical} is defined as the product
of conditional probabilities of each variable given all other variables
of a random field, \begin{equation}\label{eq_pseudo}
  PL(\boldsymbol{\theta}; \mathbf{z}) 
  = 
  \prod_{i \in \mathcal{L}} 
  \mathbb{P}(Z_{\mathbf i}= z_{\mathbf i}| \mathbf{Z}_{-{\mathbf i}} = \mathbf{z}_{-{\mathbf i}})
  =  
  \prod_{{\mathbf i}\in \mathcal{L}}
  \frac{
    e^{h_{\mathbf i}(z_{\mathbf i}|\mathbf{z})}
  }{
    \sum_{k'} e^{h_{\mathbf i}(k'|\mathbf{z})}
}.
\end{equation}

In the special case of an independent field, it is equivalent to the
likelihood function. Notice that the pseudo-likelihood function does not
depend on the intractable normalizing constant and \eqref{eq_pseudo} is
numerically equivalent to a logistic regression problem where each pixel
values corresponds to independent observations and the interacting pixel
values are covariates with coefficients corresponding to the associated
potentials.

\hypertarget{generating-mrfs-via-gibbs-sampler}{%
\paragraph{Generating MRFs via Gibbs
Sampler}\label{generating-mrfs-via-gibbs-sampler}}

While exact sampling from dependent and high-dimensional processes is a
challenging task overall, the conditional independence of MRFs
simplifies the implementation of the Gibbs Sampler algorithm
\citep{geman1984stochastic}. In the Gibbs Sampler algorithm, each pixel
value is updated conditionally to the current state of its neighbors and
a Gibbs Sampler \textbf{cycle} consists of updating each pixel exactly
one time.

To avoid introducing any kind of bias due to updates order, a random
permutation of \(\mathcal{L}\) is drawn to define the order in which
pixels are updated at each cycle. After running a suitable number of
cycles in Algorithm \ref{algo_gibbs}, the distribution of the resulting
field sampled in the process is approximately the joint distribution of
the MRF.

\begin{algorithm}[H]
\label{algo_gibbs}
\caption{Approximate Sampling algorithm
for MRFs using $T$ steps of Gibbs Sampler.}
\SetAlgoLined
    Initialize $\mathbf{z}$ with a starting configuration 
    $\mathbf{z} = \mathbf{z}^{(0)}$\;
    Initialize the iteration counter $t = 0$;
 \While{$t \leq T$}{
    Sample $\{i^{(1)}, i^{(2)}, \dots, i^{(|\mathcal{L}|)}\}$
 a random permutation of the pixel positions $\mathcal{L}$\;
    \For{$j$ in $1,\dots, |\mathcal{L}|$}{
        Update $z_{i^{(j)}}$ conditional to the rest of the field 
$\mathbf{z}_{-i^{(j)}}$ with probabilities from Equation \eqref{eq_condprob}\;
    }
  t = t + 1\;
\KwResult{output the final configuration $\mathbf{z}$.}
 }
\end{algorithm}

Sampling a field conditional to a subset of pixel values can be achieved
with the same algorithm by skipping the updates for those pixels which
are being conditioned on.

There exists faster mixing algorithms for particular cases such as
Swendsen-Wang algorithm \citep{wang1990cluster}, but they require
specific conditions from the model and/or particular implementations to
be efficient. Therefore, we keep the Gibbs Sampler as the method of
choice in this work due to its generalization ability as it only
requires computing conditional distributions, despite its slower mixing
times in some scenarios.

\hypertarget{sufficient-statistics}{%
\paragraph{Sufficient statistics}\label{sufficient-statistics}}

An important computational consequence of the model assumptions is the
fact that, in order to evaluate the probability (or likelihood) function
for a particular observed field \(\mathbf{z}\), it is not necessary to
determine the values of each pixel individually, but only the
co-occurrence counts each relative position
\({\mathbf r}\in \mathcal{R}\).

\(H(\mathbf{z},\theta)\) can be rewritten as \begin{equation}
  \label{eq_H_counts}
  H(\mathbf{z}, \theta) = 
    \sum_{{\mathbf r}\in \mathcal{R}}\sum_{a = 0}^C\sum_{b = 0}^C 
  \theta_{\mathbf r}(a,b)n_{a,b,{\mathbf r}}(\mathbf{z}),
\end{equation} where
\(n_{a,b,r}(\mathbf{z}) =  \sum_{{\mathbf i}\in \mathcal{L}} \mathbbm{1}_{(z_{\mathbf i}= a, z_{({\mathbf i}+{\mathbf r})} = b)}\)
is the count of occurrences of the pair \((a,b) \in \mathcal{Z}^2\) in
pairs of pixels with relative position \({\mathbf r}\). Therefore,
\[S_\mathcal{R}(\mathbf{z}) = \{n_{a,b,{\mathbf r}}(\mathbf{z}), a,b \in \mathcal{Z}, 
{\mathbf r}\in \mathcal{R} \}\] is a vector of sufficient statistics,
where each component \(n_{a,b,{\mathbf r}}(\mathbf{z})\) is associated
with a corresponding potential \(\theta_{\mathbf r}(a,b)\).
\citet{gimel1996texture} calls this sufficient statistic the
\emph{co-occurrence histogram}.

Parameter constraints reduce the dimension of the sufficient statistic.
Our identifiability constraint \(\theta_{\mathbf r}(0,0) = 0\) implies
that all \(n_{0,0,{\mathbf r}}(\mathbf{z})\) are excluded from
\(S_\mathcal{R}(\mathbf{z})\) and equality constraints require
aggregating (sum) co-occurrence counts to match the parameter dimension.
We shall keep the same notation for the constrained version of the
sufficient statistics \(S_{\mathcal N}(\mathbf{z})\) and potentials
\(\boldsymbol\theta\).

The main advantages of the representation with sufficient statistics are
the reduced memory usage in Monte-Carlo methods and a convenient
representation of \(H(\mathbf{z}, \boldsymbol\theta)\) with an inner
product that simplifies dealing with likelihood ratios as in
\citet{geyer1992},

\begin{equation}\label{eq_inp}
  H(\mathbf{z}, \boldsymbol\theta) =
    \langle S_\mathcal{N}(\mathbf{z}), \boldsymbol\theta \rangle.
\end{equation}

\hypertarget{sec:hmrf}{%
\subsection{Gaussian mixtures driven by Hidden MRFs}\label{sec:hmrf}}

Another class of models present in the image processing field are Hidden
Markov Random Field models (HMRFs). The hidden version considers a
latent (unobserved) process, denoted \(\mathbf{Z}\) and an observed
field, denoted \(\mathbf{Y}\), where \(\mathbf{Z}\) is distributed as a
MRF and the distribution of \(\mathbf{Y}|\mathbf{Z}\) is reasonably
simple.

In this type of modeling, \(\mathbf{Z}\) is often considered the
``true'' image and \(\mathbf{Y}\) is a noisy image. Therefore, the goal
of the analysis in this context is usually to recover the underlying
field. Note that for models like the one in this work, where
\(\mathbf{Z}\) has finite support, the hidden field defines a
segmentation of the image, making it a suitable approach for image
segmentation.

In \pkg{mrf2d}, we provide built-in tools for the case where
\(\mathbf{Y}|\mathbf{Z}\) is a finite Gaussian mixture where mixture
components are driven by the hidden field. Additional covariates can
also be included as fixed effects for the mean,

\begin{equation}
Y_{\mathbf i}|Z_{\mathbf i}= a \sim N(\mu_a + \mathbf{x}_{\mathbf i}\top \boldsymbol\beta, \sigma^2_a),
  \hspace{1cm} a = 0, 1, \dots, C.
\end{equation}

Observed values \(\mathbf{Y}\) are also considered independent given the
latent field, leading to the conditional density \begin{equation}
  f(\mathbf{y}| \mathbf{Z} = \mathbf{z}) = 
    \prod_{{\mathbf i}\in \mathcal{L}}
    \frac{1}{\sqrt{2\pi\sigma^2_{z_{\mathbf i}}}} \exp 
    \left( 
      \frac{(y_{\mathbf i}- \mu_{z_{\mathbf i}} - \mathbf{x}_{\mathbf i}^\top\boldsymbol\beta)}{2\sigma^2_{z_{\mathbf i}}}
    \right)
\end{equation} and the complete likelihood function \begin{equation}
L_{\boldsymbol{\theta}}(\boldsymbol{\beta}, \mu_a, \sigma_a, a = 0, \dots, C;
  \mathbf{y}, \mathbf{z}) = 
  \frac{1}{\zeta_\theta} e^{(H(\mathbf{z}, \theta))}
  \prod_{{\mathbf i}\in \mathcal{L}}
    \frac{1}{\sqrt{2\pi\sigma^2_{z_{\mathbf i}}}} \exp 
    \left( 
      \frac{(y_{\mathbf i}- \mu_{z_{\mathbf i}} - \mathbf{x}_{\mathbf i}^\top\boldsymbol\beta)}{2\sigma^2_{z_{\mathbf i}}}
    \right)
\end{equation}

Inference for this models involves estimating the parameters
\((\mu_k, \sigma_k)_{k = 0,1,\dots, C}\) and \(\boldsymbol{\beta}\)
associated with the Gaussian Mixture and predicting the labels of the
latent field \(\mathbf{z}\) simultaneously. Bayesian methods and the EM
algorithm are the most common approaches. The parameters of the latent
field distribution \(\boldsymbol{\theta}\) are fixed a priori and
considered tuning hyper-parameters of the algorithm.

\hypertarget{using-the-package}{%
\section{Using the Package}\label{using-the-package}}

\hypertarget{model-representation}{%
\subsection{Model representation}\label{model-representation}}

The model described in Section \ref{model-description} can be completely
characterized by three components: the random field \(\mathbf{z}\), the
interaction structure \(\mathcal{R}\) and the potentials
\(\boldsymbol\theta\). Additionally, \(\mathbf{y}\) and the mixture
parameters \((\mu_k,\sigma^2_k)_{k = 0, \dots,C}\) are included in
Hidden MRFs.

A consistent representation of each component is provided in the package
so that inputs and outputs of built-in functions, as well as methods the
user may implement, are compatible and usable in the analysis pipeline.
Representations are described in Table \ref{table_rep}.

\begin{longtable}[]{@{}ccl@{}}
\caption{Model representation summary.\label{table_rep}}\tabularnewline
\toprule
\begin{minipage}[b]{0.17\columnwidth}\centering
Model Component\strut
\end{minipage} & \begin{minipage}[b]{0.11\columnwidth}\centering
Function Argument\strut
\end{minipage} & \begin{minipage}[b]{0.64\columnwidth}\raggedright
Representation in \pkg{mrf2d}\strut
\end{minipage}\tabularnewline
\midrule
\endfirsthead
\toprule
\begin{minipage}[b]{0.17\columnwidth}\centering
Model Component\strut
\end{minipage} & \begin{minipage}[b]{0.11\columnwidth}\centering
Function Argument\strut
\end{minipage} & \begin{minipage}[b]{0.64\columnwidth}\raggedright
Representation in \pkg{mrf2d}\strut
\end{minipage}\tabularnewline
\midrule
\endhead
\begin{minipage}[t]{0.17\columnwidth}\centering
\(\mathbf{z}\): Discrete-valued field\strut
\end{minipage} & \begin{minipage}[t]{0.11\columnwidth}\centering
\code{Z}\strut
\end{minipage} & \begin{minipage}[t]{0.64\columnwidth}\raggedright
A \code{matrix} object with values in \(\{0, \dots, C\}\), where
\code{Z[w,q]} represents the pixel value in position \((w,q)\) of the
lattice. \code{NA} values are used for positions that do not belong to
\(\mathcal{L}\) when it is not a rectangular region.\strut
\end{minipage}\tabularnewline
\begin{minipage}[t]{0.17\columnwidth}\centering
\(\mathbf{y}\): Countinuous-valued field\strut
\end{minipage} & \begin{minipage}[t]{0.11\columnwidth}\centering
\code{Y}\strut
\end{minipage} & \begin{minipage}[t]{0.64\columnwidth}\raggedright
A \code{matrix} object with real values, where \code{Y[u,v]} represents
the pixel value in position \((u,v)\) of the lattice. \code{NA} values
are used for positions that do not belong to \(\mathcal{L}\) when it is
not a rectangular region.\strut
\end{minipage}\tabularnewline
\begin{minipage}[t]{0.17\columnwidth}\centering
\(\mathcal{R}\): Interaction structure\strut
\end{minipage} & \begin{minipage}[t]{0.11\columnwidth}\centering
\code{mrfi}\strut
\end{minipage} & \begin{minipage}[t]{0.64\columnwidth}\raggedright
An object of the S4 class \code{mrfi}. It can be created with the
\code{mrfi()} function.\strut
\end{minipage}\tabularnewline
\begin{minipage}[t]{0.17\columnwidth}\centering
\(\theta_{\mathbf r}(a,b)\): Array of potentials\strut
\end{minipage} & \begin{minipage}[t]{0.11\columnwidth}\centering
\code{theta}\strut
\end{minipage} & \begin{minipage}[t]{0.64\columnwidth}\raggedright
A three-dimensional \code{array} object with dimensions
\((C+1) \times (C+1) \times |\mathcal{R}|\). For a pair of values
\((a,b)\) and the \(s\)-th interacting position \({\mathbf r}_s\) of
\(\mathcal{R}\), the corresponding potential is mapped at
\code{theta[a+1, b+1, s]}.\strut
\end{minipage}\tabularnewline
\bottomrule
\end{longtable}

\hypertarget{random-fields-mathbfz-and-mathbfy.}{%
\paragraph{\texorpdfstring{Random fields \(\mathbf{z}\) and
\(\mathbf{y}\).}{Random fields \textbackslash mathbf\{z\} and \textbackslash mathbf\{y\}.}}\label{random-fields-mathbfz-and-mathbfy.}}

Realizations of a random fields \(\mathbf{z}\) and \(\mathbf{y}\) are
represented by simple \code{matrix} objects with dimension
\(N \times M\), where \(N \geq \max_{i_1} (i_1, i_2) \in \mathcal{L}\)
and \(M \geq \max_{i_2} (i_1, i_2) \in \mathcal{L}\), i.e., the maximal
coordinates. This matrix represents a rectangular set of pixels that
contains \(\mathcal{L}\). The value in row \(i_1\) and column \(i_2\)
represents the observed value of the random field in position
\((i_1,i_2)\): an integer in \(\{0, 1, \dots, C \}\) for \(\mathbf{z}\)
or a real number for \(\mathbf{y}\).

We do not require \(\mathcal{L}\) to be a complete rectangular region.
Pixels which position does not belong to \(\mathcal{L}\) are be assigned
the \code{NA} value.

Two functions are available for visualizing random fields:
\code{dplot()} and \code{cplot()}. \code{dplot()} should be used for
\textbf{d}iscrete-valued fields \(\mathbf{z}\) while \code{cplot()} is
used for \textbf{c}ontinuous-valued matrices \(\mathbf{y}\). These
functions provide an alternative to base \proglang{R} \code{image()}
function, producing elegant images in the form of \code{ggplot} objects.
The main advantage is that they allow the use of the \pkg{ggplot2}
package \citep{refggplot} to customize the image using the grammar of
graphics. Details and examples of customization of the images produced
using \pkg{ggplot2} can be found in Appendix \ref{sec:apndx_gg}.

\hypertarget{interaction-structures-mathcalr}{%
\paragraph{\texorpdfstring{Interaction Structures
\(\mathcal{R}\)}{Interaction Structures \textbackslash mathcal\{R\}}}\label{interaction-structures-mathcalr}}

Interaction structures are represented by objects of the S4 class
\code{mrfi} implemented in \pkg{mrf2d}. These objects can be created
with the \code{mrfi()} function, which has arguments \code{max_norm},
\code{norm_type} and \code{positions}. The interaction structure created
will include all relative positions which satisfy
\(||(i_1, i_2)|| \leq\) \code{max_norm} for the specified norm type.
\code{positions} can be passed as a list containing length 2 integer
vectors with relative positions to include. The function automatically
checks for repeated and opposite relative positions to ensure the
structure is valid.

\code{norm_type} options are the same as \proglang{R} built-in
\code{norm()} function, mainly, \code{"1"}, \code{"2"} and \code{"m"}
are used for \(\ell_1\), \(\ell_2\) and the maximum norm, respectively.
The default is \(\ell_1\) norm.

Some examples for creating different \(\mathcal{R}\) are detailed below.

\begin{itemize}
  \item \code{mrfi(max_norm = 1)} creates an interaction structure with all
  positions with $||(i_1, i_2)||_1 \leq 1$, which corresponds to a nearest-neighbor
  structure $\mathcal{R} = \{ (1,0), (0,1)\}$.
  \item \code{mrfi(max_norm = 0, positions = list(c(1,0), c(0,1)))} is an
  alternative way of specifying the same structure of the previous example.
  \item \code{mrfi(max_norm = 1, positions = list(c(2,0)))} results in
  the interaction structure $\mathcal{R} = \{(1,0), (0,1), (2,0) \}$.
  \item \code{mrfi(max_norm = 1, positions = list(c(-1,0)))} results in 
  $\mathcal{R} = \{(0,1), (-1,0) \}$. The norm-based and position-based
  positions had an intersection, so the redundant position $(1,0)$ was
  removed. In case of opposite directions, the \code{positions} argument is
  prioritized as it was explicitly defined by the user.
\end{itemize}

An algebra of \code{mrfi} objects is implemented for manipulating these
objects. \code{+} is used to perform union of two \code{mrfi} objects or
a \code{mrfi} object and a \code{numeric} vector with 2 integers can be
used to add a single interacting position to an existing \code{mrfi}
object. Similarly, the \code{-} operator can be used to perform set
difference between two \code{mrfi} objects or to remove a single
position if a vector with 2 integers is used in the right-hand-side.

Additionally, conversion of \code{mrfi} objects to \code{list} is
implemented in the \code{as.list()} method. Subsetting methods are also
available with the \code{"[]"} and \code{"[[]]"} operators. These
methods are particularly important for model selection algorithms, as
many distinct sparse interaction structures can be obtained by using
different subsets of a large reference base structure.

A \code{plot} method is available for \code{mrfi} objects. The code
chunk below exemplifies the usage of plotting functions and manipulation
of \code{mrfi} objects. The resulting plots are presented in Figure
\ref{fig:example_mrfi_plot}. The black square represents the origin
position \((0,0)\), positions included in the interaction structure
\(\mathcal{R}\) are represented by the dark-gray squares with black
borders, while their opposite directions are the light-gray squares.

\begin{CodeChunk}

\begin{CodeInput}
R> plot(mrfi(max_norm = 1))
R> plot(mrfi(max_norm = 2, norm_type = "m") + c(4,0))
R> plot(mrfi(4) - mrfi(2))
R> plot(mrfi(6, norm_type = "m")[c(1,2,6,9,19,41)])
\end{CodeInput}
\begin{figure}[h]

{\centering \includegraphics[width=0.24\linewidth]{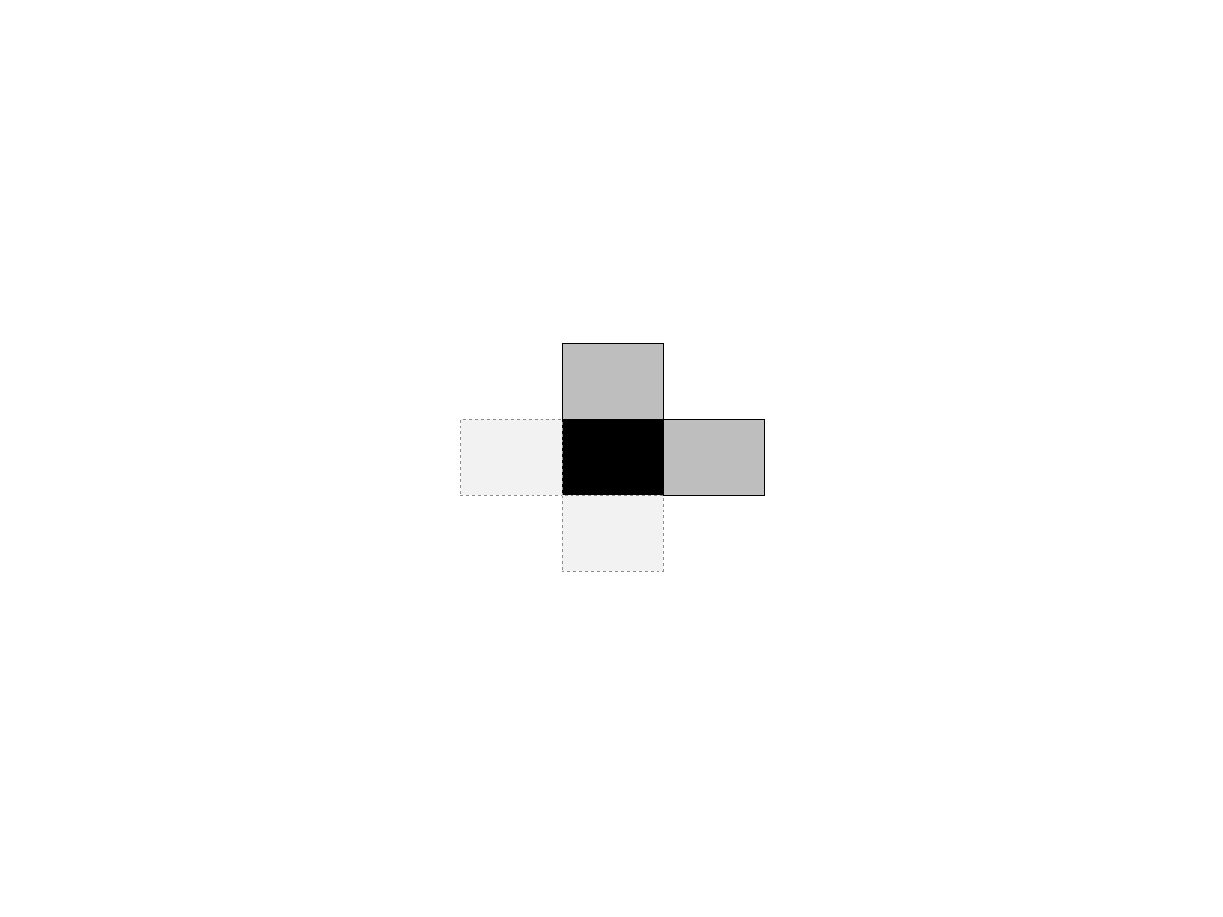} \includegraphics[width=0.24\linewidth]{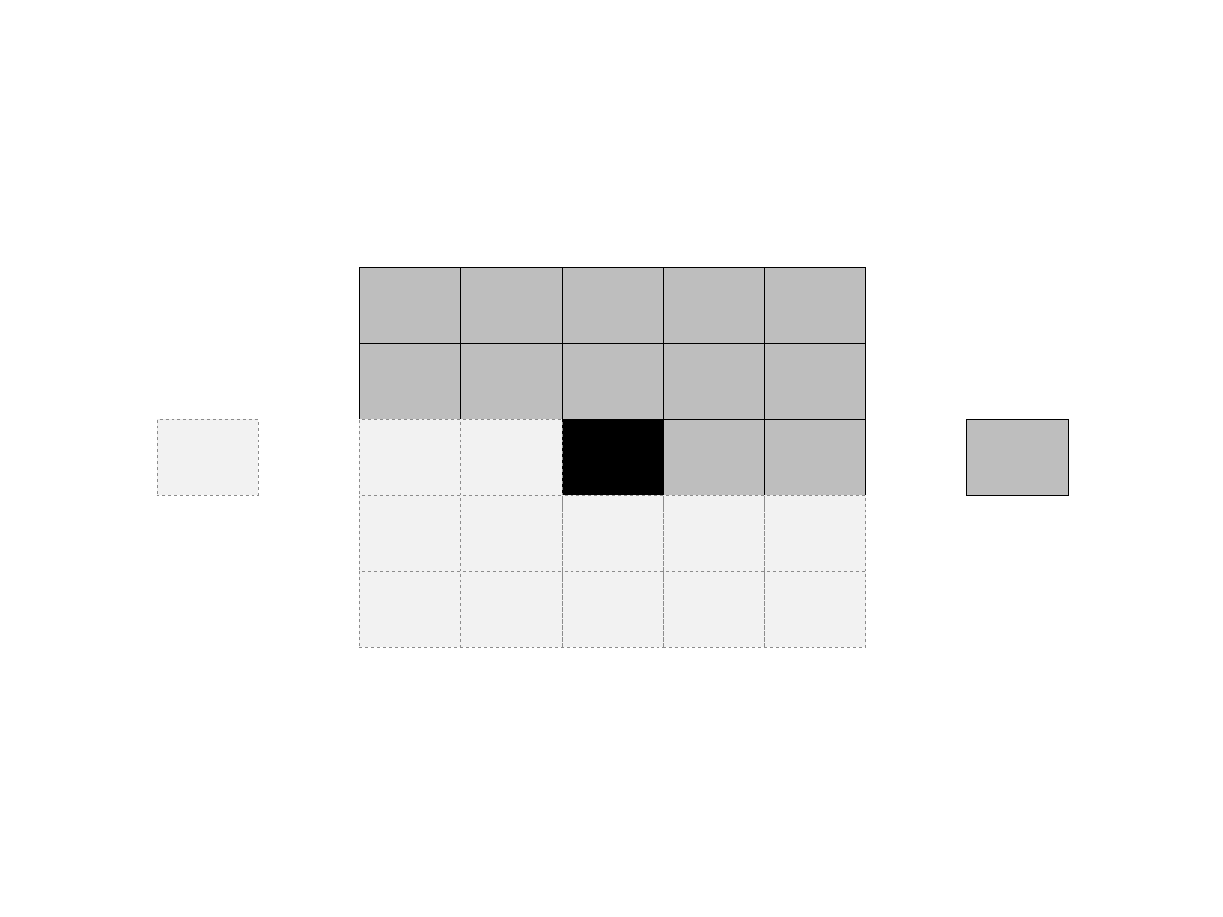} \includegraphics[width=0.24\linewidth]{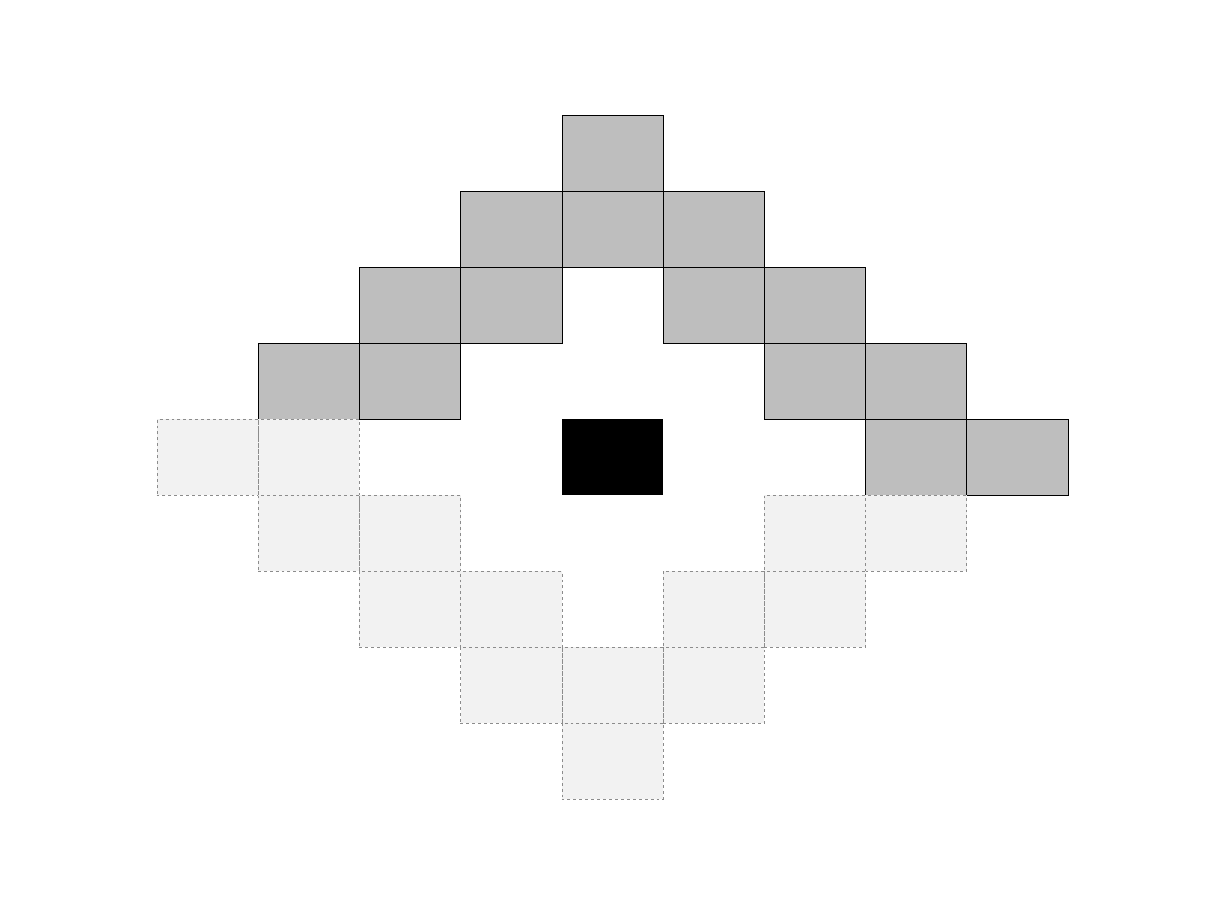} \includegraphics[width=0.24\linewidth]{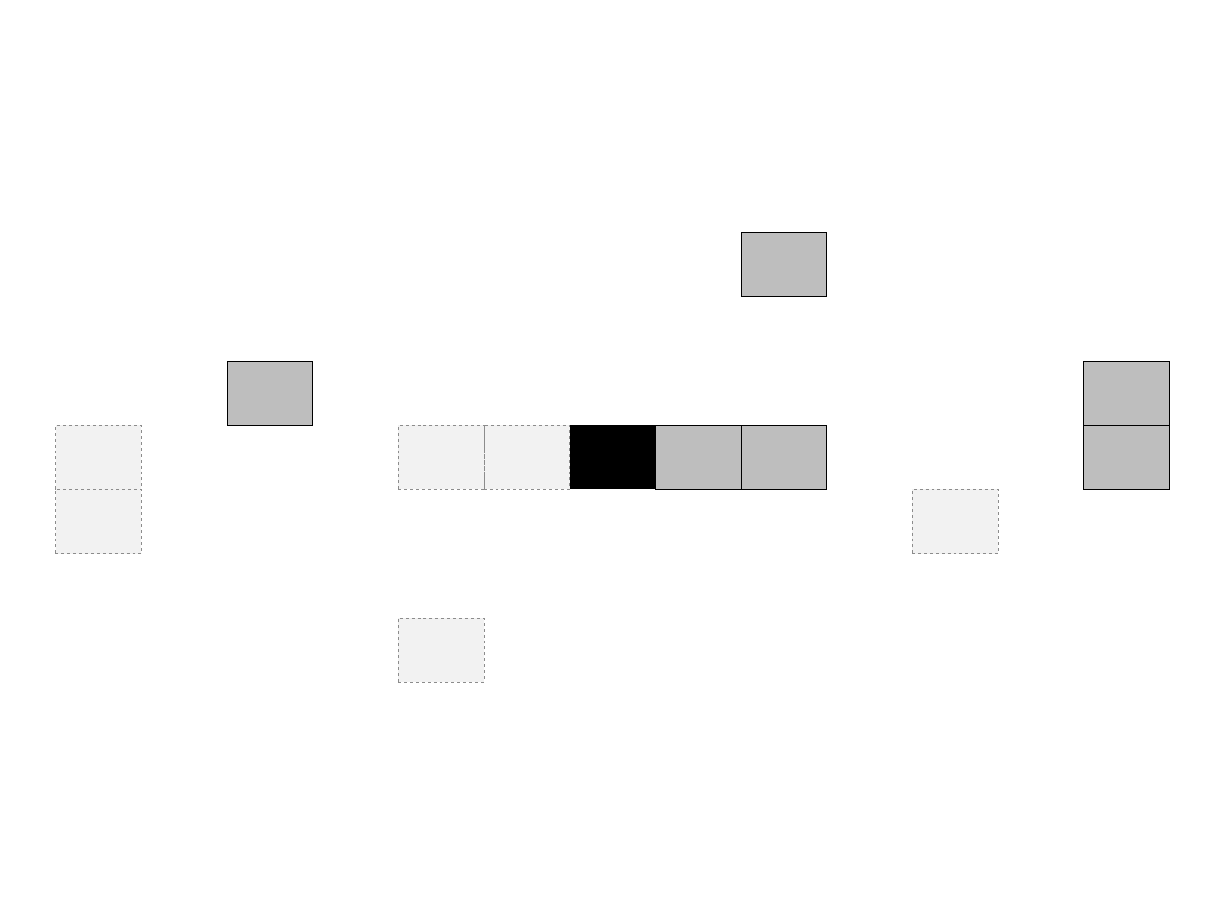} 

}

\caption[Examples of interaction structures $\mathcal{R}$ created and their visualization]{Examples of interaction structures $\mathcal{R}$ created and their visualization.}\label{fig:example_mrfi_plot}
\end{figure}
\end{CodeChunk}

\hypertarget{potentials-array-boldsymboltheta}{%
\paragraph{\texorpdfstring{Potentials array
\(\boldsymbol\theta\)}{Potentials array \textbackslash boldsymbol\textbackslash theta}}\label{potentials-array-boldsymboltheta}}

The collection of potentials, \(\theta_{\mathbf r}(a,b)\), is
represented by an \code{array} object with dimensions
\((C+1) \times (C+1) \times |\mathcal{R}|\). Rows and columns are used
to map \(a\) and \(b\), respectively, while slices are used to map
relative positions \({\mathbf r}\). A set of potentials
\(\{\theta_{\mathbf r}(a,b), a,b \in \mathcal{Z}, {\mathbf r}\in \mathcal{R} \}\)
is always related to an interaction structure
\(\mathcal{R} = \{ {\mathbf r}_1, {\mathbf r}_2, \dots, {\mathbf r}_{|\mathcal{R}|} \}\).
The \(i\)-th slice maps the \(i\)-th relative position of
\(\mathcal{R}\), \(r_i\).

\begin{figure}[ht]
\centering
\includegraphics[width=0.6\linewidth]{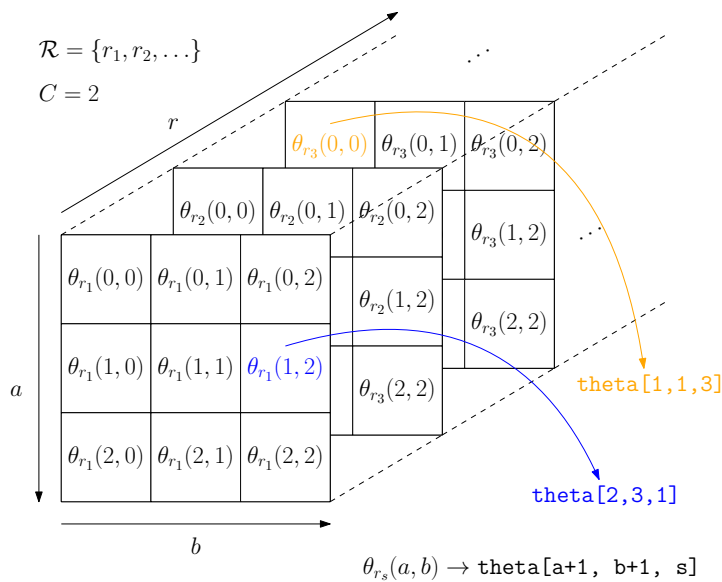}
\caption{Example of array representation of potentials with $C = 2$.}
\label{array_rep}
\end{figure}

An important detail is that array indices in \proglang{R} starts at
\code{1}, while we consider our set of possible values
\(\mathcal{Z}  = \{0, 1 , \dots, C \}\), therefore we need to shift
\(a\) and \(b\) one position when accessing their value in the
\proglang{R} array. Figure \ref{array_rep} illustrates how potentials
can be represented as an array in \proglang{R} in the \(C=2\) case. Two
elements are highlighted and the associated indices used to access them
are shown as examples.

\hypertarget{sec:apndx_par}{%
\subsection{Parameter restriction families}\label{sec:apndx_par}}

Parameter restrictions play an important role in the inference process
of our Markov Random Field models. \pkg{mrf2d} functions support 5
families of parameter restrictions for the array of potentials to be
considered in inference algorithms. They are specified by the
\code{family} argument of functions to ensure the resulting output array
(\code{theta}) respects those constraints. A brief description of each
interaction structure is given next. Table \ref{tbl_families} presents
the mathematical definitions, number of free parameters and an example
of a slice of the array of potentials for the case with \(C = 2\) in
each family.

\begin{table}[ht]
\caption{Description of parameter restriction families.}
\label{tbl_families}
\centering
\begin{tabular}{c | c | c | c}
\toprule
Family & Restriction  & Free parameters  & Example slice\\ \midrule
\code{"onepar"} & $\theta_\mathbf{r}(a,b) = \phi \mathbbm{1}_{(a\neq b)}$ &
$1$ &
$\begin{bmatrix}
0 & \phi & \phi \\
\phi & 0 & \phi \\
\phi & \phi & 0
\end{bmatrix}$ \\

\code{"oneeach"} & $\theta_\mathbf{r}(a,b) = \phi_{\mathbf r}\mathbbm{1}_{(a\neq b)}$ &
$|\mathcal{R}|$& 
$\begin{bmatrix}
0 & \phi_{\mathbf r}& \phi_{\mathbf r}\\
\phi_{\mathbf r}& 0 & \phi_{\mathbf r}\\
\phi_{\mathbf r}& \phi_{\mathbf r}& 0
\end{bmatrix}$ \\

\code{"absdif"} & 
$\theta_{\mathbf r}(a,b) = \sum_{d = 1}^C \phi_{\mathbf{r}, d} \mathbbm{1}_{(|b-a| = d)}$ &
$|\mathcal{R}|C$ & 
$\begin{bmatrix}
0 & \phi_{{\mathbf r},1} & \phi_{{\mathbf r},2} \\
\phi_{{\mathbf r},1} & 0 & \phi_{{\mathbf r},1} \\
\phi_{{\mathbf r},2} & \phi_{{\mathbf r},1} & 0
\end{bmatrix}$ \\

\code{"dif"} & $\theta_{\mathbf r}(a,b) = \sum_{d = -C, d \neq 0}^C \phi_{{\mathbf r}, d} \mathbbm{1}_{(b-a = d)}$ & $|\mathcal{R}|2C$ &
$\begin{bmatrix}
0 & \phi_{{\mathbf r},1} & \phi_{{\mathbf r},2} \\
\phi_{{\mathbf r},-1} & 0 & \phi_{{\mathbf r},1} \\
\phi_{{\mathbf r},-2} & \phi_{{\mathbf r},-1} & 0
\end{bmatrix}$ \\

\code{"free"} & $\theta_{\mathbf r}(0,0) = 0$ & $|\mathcal{R}|(C^2 - 1)$ &
$\begin{bmatrix}
0 & \phi_{{\mathbf r},0,1} & \phi_{{\mathbf r},0,2} \\
\phi_{{\mathbf r},1,0} & \phi_{{\mathbf r},1,1} & \phi_{{\mathbf r},1,2} \\
\phi_{{\mathbf r},2,0} & \phi_{{\mathbf r},2,1} & \phi_{{\mathbf r},2,2}
\end{bmatrix}$ \\ \bottomrule
\end{tabular}
\end{table}

\begin{description}

\item[\code{"onepar"}] A single-parameter ($\phi$) model, where interactions
depend only on the fact that values are equal or different, regardless of their
relative position. This restriction corresponds to the classical Ising and
Potts model.

\item[\code{"oneeach"}] The same interaction type as \code{"onepar"}, but
allowing different values $\phi_\mathbf{r}$ for different interacting positions
$\mathbf{r} \in \mathcal{R}$.

\item[\code{"absdif"}] For each $\mathbf{r} \in \mathcal{R}$, the potentials
$\theta_\mathbf{r}(a,b)$ are equal when the absolute differences of their pixel
values $d = |b-a|$ are the same.
Note that \code{"absdif"} is equivalent to \code{"oneeach"} when $C = 1$.

\item[\code{"dif"}] 
Generalizes the \code{"absdif"} family allowing opposite signal differences
to have different interactions.

\item[\code{"free"}]
No restrictions, except for the identifiability constraint $\theta_{\mathbf r}(0,0) = 0$.

\end{description}

Families \code{"dif"} and \code{"absdif"} should only be used when pixel
values are actually quantities and differences are well-defined, for
example, in grayscale images with few levels, for example in
\citet{gimel1996texture}. If relabeling the values does not change the
interpretation of the problem, then these restrictions are probably not
suitable.

The function \code{smr_array(theta, family)} can be used to transform a
parameter array into a vector of appropriate length containing only the
free parameters corresponding to the provided array (\code{theta}) and
the restriction \code{family}. The opposite operation is also available
as the \code{expand_array(theta_vec, family, mrfi, C)} function. This
transformations are particularly useful for optimization problems as
most functions, for example, \proglang{R} built-in \code{optim} function
requires a vector of parameters and for storing multiple vectors, for
example in Monte-Carlo methods, using a simpler and less memory
consuming structure.

\hypertarget{random-field-sampler}{%
\subsection{Random field sampler}\label{random-field-sampler}}

Being able to sample observations of Markov random fields is a key
component of many inference methods that aim to avoid the intractable
normalizing. In \pkg{mrf2d}, a complete and efficient routine to sample
fields using the Gibbs Sampler algorithm described in Algorithm
\ref{algo_gibbs} is provided by the \code{rmrf2d()} function. Its
arguments are:

\begin{itemize}
  \item \code{init_Z}: The initial field configuration, or a length-2 vector
  with the dimensions of the field to be sampled. If the dimensions are provided,
  the initial configuration is randomly sampled from independent discrete uniform
  distributions.
  \item \code{mrfi}: A \code{mrfi} object representing the interaction structure
  $\mathcal{R}$.
  \item \code{theta}: An \code{array} of potentials.
  \item \code{cycles}: The number of Gibbs Sampler cycles.
  \item \code{sub_region}: Optional argument used for non-rectangular images when \code{init_Z} is
  a vector with the dimensions. A \code{logical} matrix with the same dimensions
  as specified in \code{init_Z}. Pixels with \code{FALSE} value are not included
  in the image.
  \item \code{fixed_region}: Optional. A matrix with \code{logical} values. 
  Pixel positions with \code{TRUE} value are conditioned on their initial
  configuration (\code{init_Z}) value and are not updated.
\end{itemize}

We illustrate the use of the sampling function below on two
\(200 \times 200\) fields: one without conditioning on any pixel
(nothing specified in \code{fixed_region}) and one conditioned on the
border values set as \(0\). The resulting images are presented in Figure
\ref{fig:sampler_examples}.

\begin{CodeChunk}

\begin{CodeInput}
R> th <- expand_array(-1, family = "onepar", mrfi(1), C = 1)
R> 
R> # Sample a field
R> z_sample <- rmrf2d(init_Z = c(200,200), mrfi = mrfi(1), theta = th)
R> 
R> # Sample a field conditional to the border values being 0
R> ## Define the logical matrix for the fixed region.
R> border <- matrix(FALSE, nrow = 100, ncol = 100) 
R> border[1,] <- border[100,] <- border[,1] <- border[,100] <- TRUE 
R> ## Define the initial field
R> initial <- matrix(sample(0:1, 100*100, replace = TRUE), 
+   nrow = 100, ncol = 100)
R> initial[border] <- 0
R> z_border <- rmrf2d(initial, mrfi = mrfi(1), theta = th, 
+   fixed_region = border)
\end{CodeInput}
\end{CodeChunk}

\begin{CodeChunk}
\begin{figure}

{\centering \includegraphics[width=0.49\linewidth]{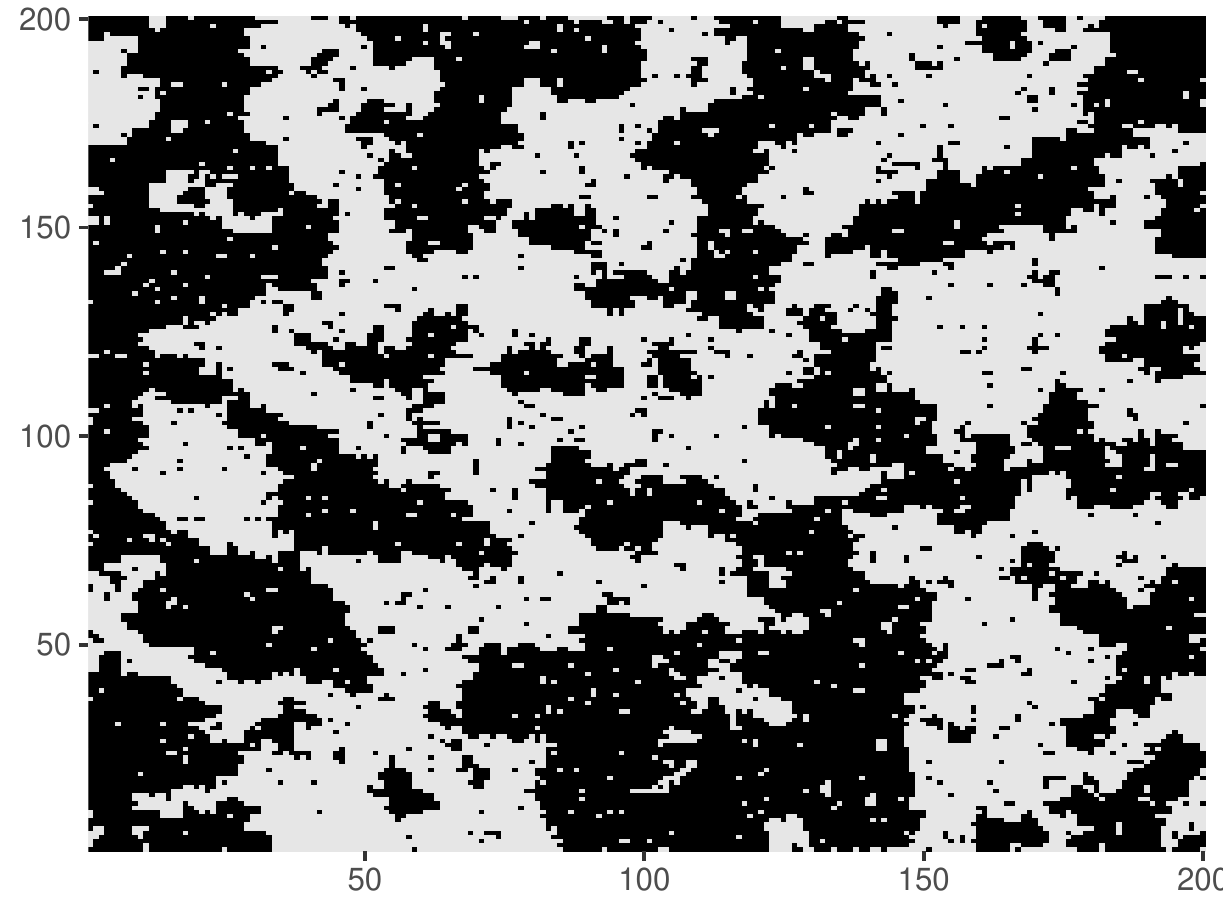} \includegraphics[width=0.49\linewidth]{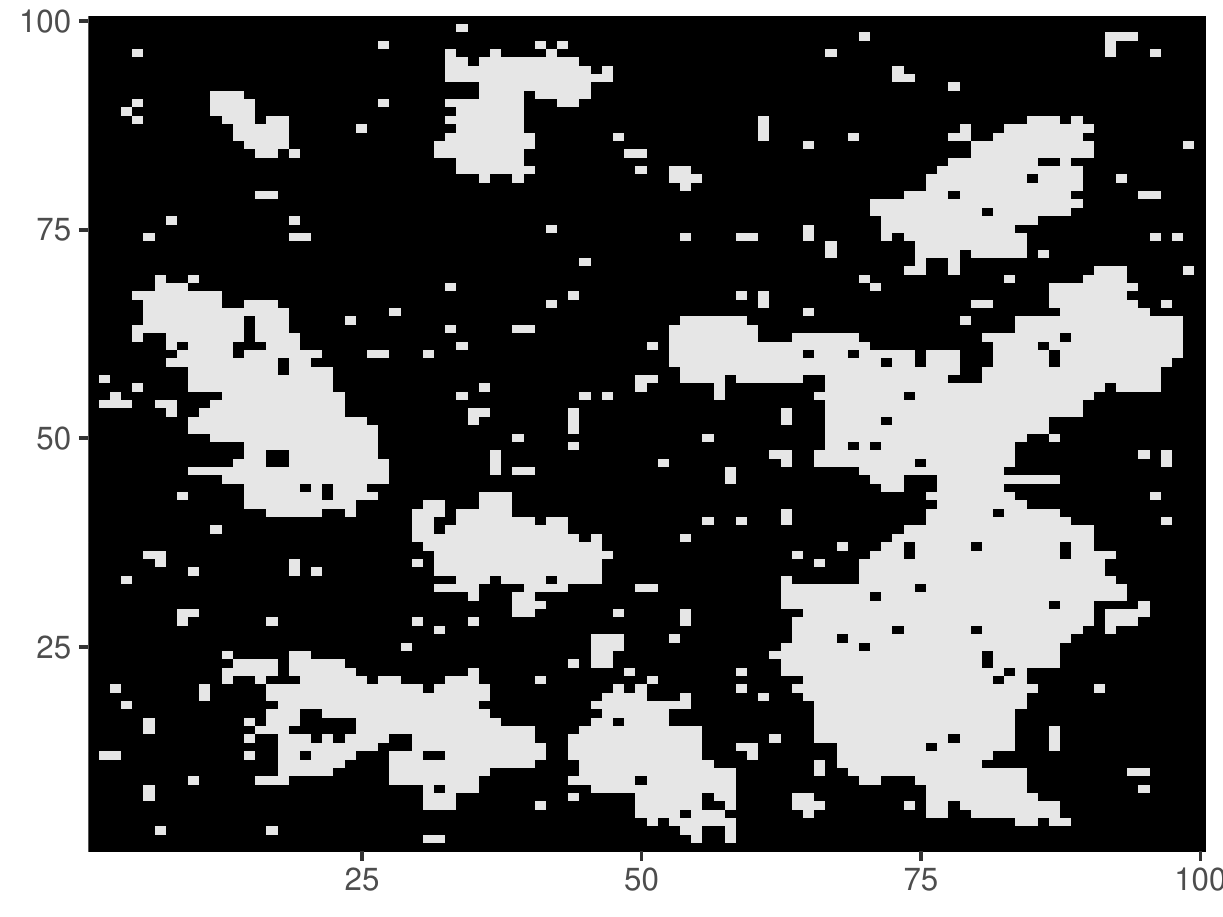} 

}

\caption[Simulated random fields from a nearest-neighbor structure]{Simulated random fields from a nearest-neighbor structure. Left: no boundary conditions. Right: conditional to all border values being 0 (black).}\label{fig:sampler_examples}
\end{figure}
\end{CodeChunk}

Fields with non-rectangular can be sampled either by passing a
non-rectangular field as the \code{init_Z} argument or by using the
\code{sub_region} argument and specifying the dimensions of the sampled
field.

Another important feature is conditioning on a subset of pixel values.
There are many situations where keeping a subset of pixels fixed during
the sampling process can be useful, for example, filling a region of
missing pixel values via simulation, defining boundary conditions (our
model corresponds to a free boundary condition, but other types such as
fixed or periodic boundary can be sampled with proper manipulation of
the initial configuration and conditioning region) or performing
block-wise updates of the data using conditionally independent blocks
(for parallelization of algorithms).

Since the Gibbs Sampler algorithm updates each pixel value multiple
times, performance is one of our main implementation concerns. To
improve the performance and speed up computations considerably, the
internals of the sampling function, as well as most other
computationally intensive functions are written in \proglang{C++} with
the use of \pkg{Rcpp}\citep{ref_rcpp} and
\pkg{RcppArmadillo}\citep{ref_armadillo} packages.

\hypertarget{statistical-inference-in-mrf2d}{%
\subsection{Statistical Inference in
mrf2d}\label{statistical-inference-in-mrf2d}}

Inference methods for MRF models are diverse and their suitability
highly depend on the type of data being analyzed. The framework provided
by \pkg{mrf2d} can be used to implement all sorts of algorithms that are
built from a common stack of components: simulation, conditional
probabilities and sufficient statistics. It also provides complete
built-in routines for some estimation algorithms.

Table \ref{tbl_functions} presents a list of functions available in the
package that can be used to construct inference algorithms, as well as
built-in functions for parameter estimation for MRF and for the Hidden
MRF models defined in Section \ref{sec:hmrf} that we describe next. The
built-in inference functions return objects of class \code{mrfout} (MRF
data) or \code{hmrfout} (Hidden MRF models), which contains the
information about the fitted model, as well as \code{summary} and
\code{plot} methods associated for interpretation of the results.

\begin{table}[ht]
\caption{List of available functions used for inference in \pkg{mrf2d} with a brief description of each one.}
\label{tbl_functions}
\centering
\begin{tabular}{p{0.2\columnwidth} p{0.8\columnwidth}}
\toprule
\multicolumn{1}{l}{Function} & \multicolumn{1}{l}{Use} \\
\midrule
\multicolumn{2}{c}{Miscellaneous}\\
\midrule
\code{rmrf2d} & Generates samples of a MRF via Gibbs Sampler. Used for Monte-Carlo based methods.\\
\code{cp_mrf2d} & Computes the conditional probabilities for a pixel position given its neighbors.\\
\code{pl_mrf2d} & Computes pseudo-likelihood value for an observed field considering interaction structure \code{mrfi} and array of potentials \code{theta}. \\
\code{cohist} & Creates the co-occurrence histogram of an observed given an interaction structure. Can be converted to a vector of sufficient statistics given a restriction family with the \code{smr_stat} function.\\
\code{smr_array} and \code{expand_array} & Conversions between array and vector representation of potentials given a parameter restriction family.\\
\midrule
\multicolumn{2}{c}{Built-in inference algorithms}\\
\midrule
\code{fit_pl} & Estimates the parameter array given an observed field via pseudo-likelihood optimization. Returns a \code{mrfout} object.\\
\code{fit_sa} & Estimates the parameter array given an observed field via Stochastic Approximation algorithm. Returns a \code{mrfout} object.\\
\code{fit_ghm} & Fits a Gaussian Mixture driven by a given Hidden MRF model using the EM algorithm from \citet{zhang2001segmentation}. \code{polynomial_2d} and \code{fourier_2d} can be used to create polynomial and 2-dimensional Fourier basis functions, respectively, to be used as a fixed effect. Returns a \code{hmrfout} object.\\
\bottomrule
\end{tabular}
\end{table}

\hypertarget{maximum-pseudo-likelihood-estimation}{%
\paragraph{Maximum pseudo-likelihood
estimation}\label{maximum-pseudo-likelihood-estimation}}

The pseudo-likelihood function in (\ref{eq_pseudo}) can be evaluated
efficiently because it does not depend on the intractable normalizing
constant. A common estimation procedure for intractable likelihood
problems is optimizing the pseudo-likelihood with respect to the
parameters.

\begin{equation}
\hat{\theta}_{PL} = \arg \max_\theta PL(\theta; \mathbf{z}).
\end{equation}

The function \code{fit_pl()} from \pkg{mrf2d} implements an optimization
of the pseudo-likelihood function using \proglang{R} built-in
\code{optim()} function. It handles the conversions between array and
vector representation of potentials automatically, respecting the
restriction family selected and returns the estimated array of
potentials and maximum value of the pseudo-likelihood in logarithmic
scale. The arguments of \code{fit_pl} are:

\begin{itemize}
  \item \code{Z}: The observed random field $\mathbf{z}$.
  \item \code{mrfi}: A \code{mrfi} representing an interaction structure $\mathcal{R}$.
  \item \code{family}: A parameter restriction family.
  \item \code{init}: An array with the initial configuration used in the optimization.
    \code{0} can be used to start from the independent model.
  \item \code{optim_args}: A named \code{list} with additional arguments passed
    to the \code{optim()} function call.
\end{itemize}

\hypertarget{stochastic-approximation-algorithm}{%
\paragraph{Stochastic Approximation
algorithm}\label{stochastic-approximation-algorithm}}

Given an observed field \(\mathbf{z}^{(0)}\), the Stochastic
Approximation algorithm \citep{ref_sa} seeks to create a Markov Chain of
parameter vectors \(\{\boldsymbol\theta^{(t)}\}_{t \geq 1}\) that
converges to the maximum likelihood estimate of \(\boldsymbol\theta\),
which is the solution of the zero gradient condition
\(\mathbb{E}_{\boldsymbol\theta}(S_\mathcal{R}(\mathbf{Z})) = S_\mathcal{R}(\mathbf{z}^{(0)})\),
derived from (\ref{eq_inp}).

The algorithm is defined by the recurrence \begin{equation}\label{eq_sa}
\boldsymbol\theta^{(t+1)} = \boldsymbol\theta^{(t)} + \gamma^{(t)}
(S_\mathcal{R}(\mathbf{z}^{(0)}) - S_\mathcal{R}(\mathbf{z}^{(t)})),
\end{equation} where \(\mathbf{z}^{(t)}\) is a field sampled using
\(\boldsymbol\theta^{(t)}\) and \(\gamma^{(t)}\) is a sequence of
positive constants that satisfies
\(\sum_{t = 1}^{\infty} \gamma^{(t)} = \infty\) and
\(\sum_{t = 1}^\infty \left(\gamma^{(t)}\right)^2 < \infty\).

Stochastic Approximation is implemented in \pkg{mrf2d} as the
\code{fit_sa()} function. It samples \(\mathbf{z^{(t)}}\) via Gibbs
Sampler considering the previous field \(\mathbf{z}^{(t-1)}\) as the
initial configuration. Periodically, the field samples are refreshed,
starting from an independent discrete uniform distribution and running a
greater number of Gibbs Sampler cycles, what prevents the algorithms
from getting stuck in problematic field samples. Its arguments are:

\begin{itemize}
    \item \code{Z} The observed field $\mathbf{z}^{(0)}$.
  \item \code{mrfi} The interaction structure $\mathcal{R}$.
  \item \code{family} The family of parameter restrictions considered when
converting the potentials array to a vector.
  \item \code{gamma_seq} A sequence of step size values to be used as 
$\gamma^{(t)}$. These values are divided by the number of pixels $|\mathcal{L}|$
internally to be invariant with respect to the image size. The typical sequence
recommended is \code{seq(from = M, to = 0, length.out = B)}, with \code{M} ranging
from $0.5$ to $2$ and large number of iterations (\code{B}).
  \item \code{init} The initial array of parameters or the value \code{0} to
start from the independent model.
  \item \code{cycles} Number of Gibbs Sampler ran between iterations.
  \item \code{refresh_each} Restarts the sample $\mathbf{z}^{(t)}$ from a random
configuration each \code{refresh_each} iterations.
  \item \code{refresh_cycles} When a refresh happens, how many Gibbs Sampler cycles
are ran in the current parameter configuration.
\end{itemize}

Among the elements of the returned \code{mrfout} object, \code{theta}
contains the estimated potential array and \code{metrics} a data frame
with the Euclidean distances between \(S_\mathcal{R}(\mathbf{z}^{(0)})\)
and \(S_{\mathcal{R}}(\mathbf{z}^{(t)})\) for each iteration. This
sequence of distances is used to monitor the convergence of the
algorithm as a form of diagnostics analysis.

\hypertarget{em-algorithm-for-hmrf-models}{%
\paragraph{EM algorithm for HMRF
models}\label{em-algorithm-for-hmrf-models}}

Gaussian Mixtures driven by Hidden MRFs can be fitted in \pkg{mrf2d}
with an extension of the EM algorithm from \citet{zhang2001segmentation}
to include a fixed effect \citep[see][ for
details]{freguglia2019hidden}. The probabilities computed for the latent
label of each pixel in the E-step are conditioned on the global maximum
probability configuration of its neighbors, obtained via Iterated
Conditional Modes (ICM) algorithm at each iteration
\citep{besag1986statistical}.

The complete algorithm is available in the \code{fit_ghm} function. Its
main arguments are

\begin{itemize}
  \item \code{Y}: The observed continuous-valued field $\mathbf{y}$.
  \item \code{mrfi}: Interaction structure of the latent field $\mathcal{R}$.
  \item \code{theta}: The array of potentials that defines the latent field distribution.
  \item \code{fixed_fn}: A \code{list} of functions of pixel positions 
  $f(i_1, i_2)$ to be used as fixed effect. Constructors for 2-dimensional 
  polynomials and Fourier basis are available in the functions 
\code{polynomial_2d} and \code{fourier_2d}, respectively.
  \item \code{equal_vars}: A \code{logical} value indicating if mixture components 
  are forced to have equal variances.
  \item \code{init_mus} and \code{init_sigmas}: Optional initial values of 
  $(\mu_a, \sigma_a)_{a = 0, \dots, C}$. If none is passed, an independent
  Gaussian mixture is fitted with initial values based on quantiles and the
  estimates of this fitted model are used as (often good) starting values in
  the main procedure.
  \item \code{maxiter}: Maximum number of iterations before stopping.
  \item \code{max_dist}: Defines a stopping condition for the EM algorithm.
  For consecutive iterations $t$ and $t+1$, the absolute difference in each
  parameter, $|\mu^{(t)}_k - \mu^{(t+1)}_k|$ and $|\sigma^{(t)}_k - \sigma^{(t+1)}_k|$
  are computed for $k = 0, 1, \dots, C$. The algorithms stops if all differences
  are less than \code{max_dist}.
  \item \code{icm_cycles}: Number of cycles of Iterated Conditional Modes 
  algorithm executed in each iteration.
\end{itemize}

\code{fit_ghm()} returns a \code{hmrfout} object containing a
\code{data.frame} with estimates of the mixture parameters
\(\{(\hat\mu_a, \hat\sigma_a), a = 0, 1 \dots, C \}\) named \code{par},
the highest probability configuration of the latent field computed via
ICM algorithm \(\mathbf{\hat z}\) in \code{Z_pred}, a matrix with the
estimated fixed effects
\(\left( \mathbf{x}_{\mathbf i}^\top \hat\beta\right)\) for each pixel
in \code{fixed} and a matrix with the predicted mean for each pixel
\(\left( \mathbf{x}_{\mathbf i}^\top \hat\beta + \hat\mu_{\hat z_{\mathbf i}}\right)\)
in \code{predicted}.

\hypertarget{sec:examples}{%
\section{Data analysis using mrf2d}\label{sec:examples}}

\hypertarget{example-1-a-binary-image-with-texture-like-pattern}{%
\subsection{Example 1: A binary image with texture-like
pattern}\label{example-1-a-binary-image-with-texture-like-pattern}}

\hypertarget{description}{%
\paragraph{Description}\label{description}}

To illustrate the usage of \pkg{mrf2d} for finite-valued images, we use
the object \code{field1} available in the package, which contains a
binary field with anisotropic pattern as seen in Figure
\ref{fig:view_field1}. It is a synthetic texture image of the same type
as the binary texture data presented in \citet{cross1983markov}. The
data can be loaded and viewed using the code chunk below.

\begin{CodeChunk}

\begin{CodeInput}
R> data("field1", package = "mrf2d")
R> dplot(field1, legend = TRUE)
\end{CodeInput}
\begin{figure}[ht]

{\centering \includegraphics[width=0.49\linewidth]{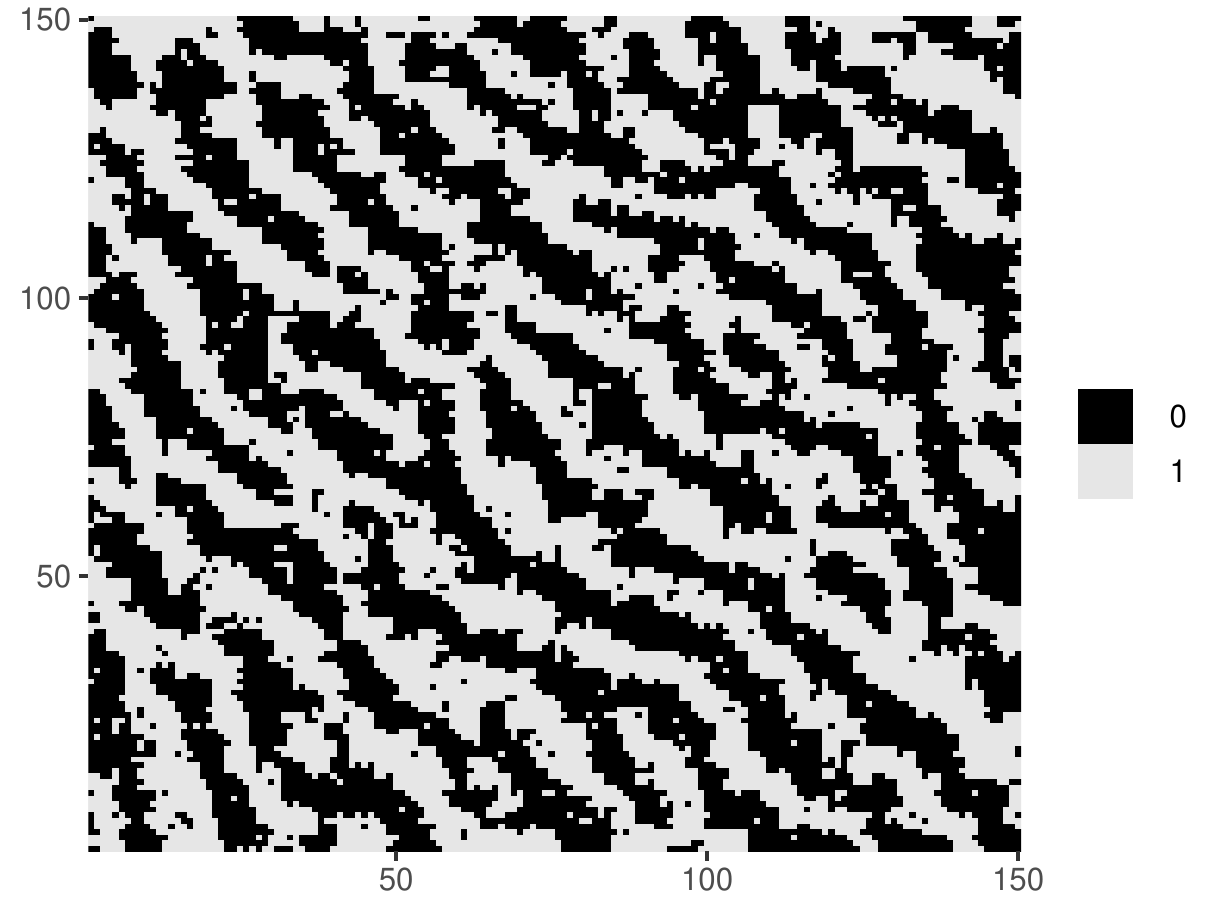} 

}

\caption[Visualization of \code{field1}]{Visualization of \code{field1}.}\label{fig:view_field1}
\end{figure}
\end{CodeChunk}

Our goal is to fit a MRF model to this data and sample images from the
fitted model to evaluate if the patterns achieved in the generated data
are similar to the original data. This is the typical setup of a texture
synthesis problem with finite-valued images.

This analysis involves three main stages: Specifying the model
(interaction structure and parameter restrictions), estimating the
parameters and evaluating the fitted model.

\hypertarget{specifying-mathcalr-and-parameter-family}{%
\paragraph{\texorpdfstring{Specifying \(\mathcal{R}\) and parameter
family}{Specifying \textbackslash mathcal\{R\} and parameter family}}\label{specifying-mathcalr-and-parameter-family}}

Model selection under intractability is a challenging problem because
most algorithms require comparing (maximum) likelihood functions for
different models, what cannot be done exactly and/or have a high
computational cost for MRF models.

The main routes in the model specification stage are: using prior
information of the data or problem to select what type of restrictions
and interaction structure are best suited, using the most general model
(e.g., no restrictions and a complex interaction structure) as in
\citet{freguglia2019hidden} or using some estimation technique.

This image presents a diagonal pattern what indicates a nearest-neighbor
interaction structure may not be appropriate to capture all the
dependence present in the field. We first choose what kind of parameter
restriction family will be considered by checking that relabeling the
values \(\mathcal{Z}\) does not change the patterns in the image
indicating symmetric potentials should be suited for this image, and
there is a clear difference in the interactions when considering pixels
in different directions, what indicates we need different types of
interaction for different relative positions. These characteristics
match the \code{"oneeach"} family that will be used in this example.

For estimating the set interacting positions \(\mathcal{R}\), we use use
a naive algorithm which consists of performing 300 steps of Stochastic
Approximation considering a large set of candidate interacting positions
(all positions with maximum norm less or equal \(6\), \(84\) positions
total) and then select the positions which absolute value of the
associated potential is higher than a threshold value. This is a
strategy similar to the heuristic search algorithm from
\citet{gimel1996texture}. Stochastic approximation was preferred over
maximum pseudo-likelihood, for example, because it is computationally
more suited for high-dimensional situations and we are not requiring a
very accurate estimation at this point, so we can use a relatively lower
number of steps.

The code below implements this naive interaction selection algorithm in
a few lines using the tools available in \pkg{mrf2d} considering a
threshold value of \(0.10\).

\begin{CodeChunk}

\begin{CodeInput}
R>  # Define a large set of interacting positions
R> candidates <- mrfi(6, norm_type = "m")
R> 
R>  # Stochastic approximations for the candidate set
R> set.seed(1)
R> complete_sa <- fit_sa(field1, candidates, family = "oneeach", 
+   gamma_seq = seq(from = 1, to = 0, length.out = 300),
+   cycles = 2, refresh_each = 301)
\end{CodeInput}
\end{CodeChunk}

\begin{CodeChunk}

\begin{CodeInput}
R> plot(complete_sa)
\end{CodeInput}
\begin{figure}

{\centering \includegraphics[width=0.66\linewidth]{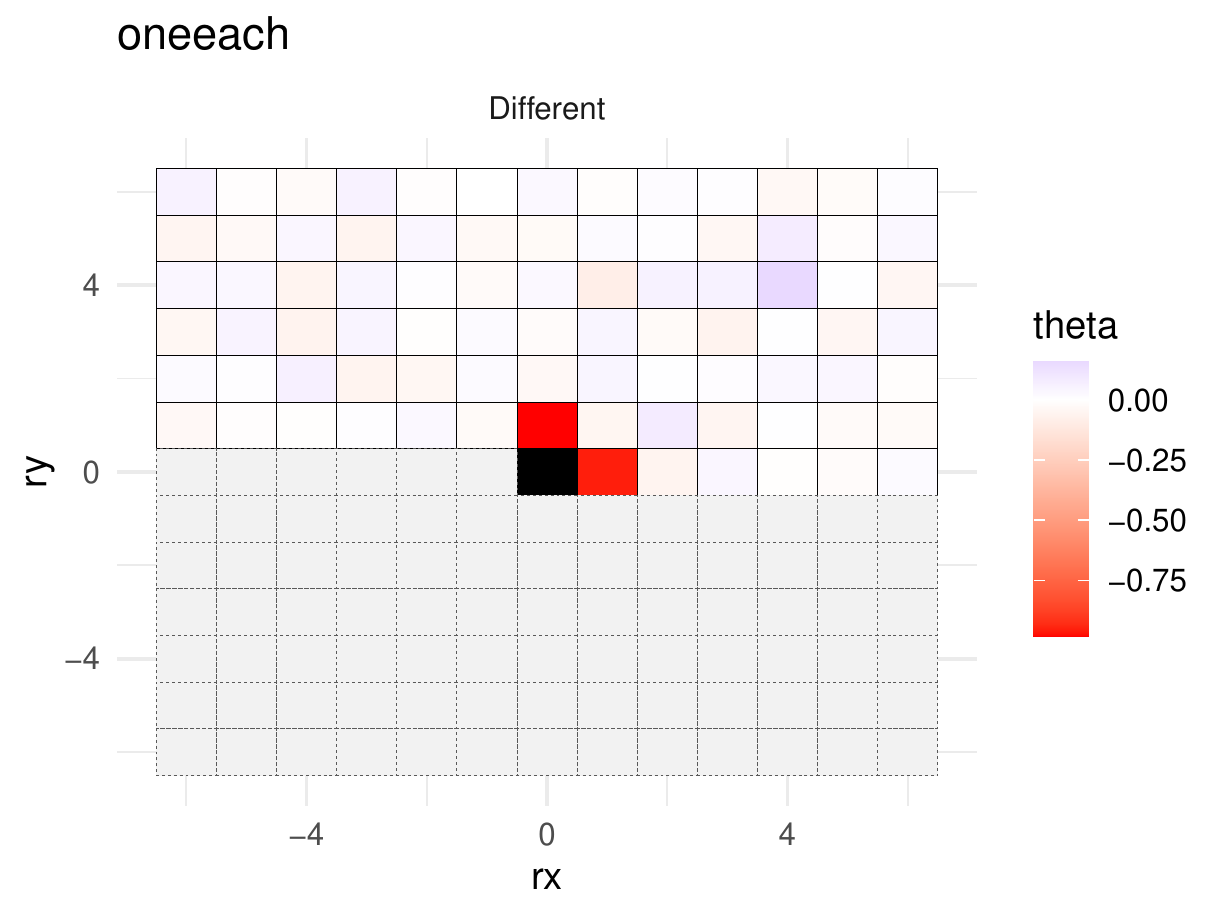} 

}

\caption[Plot of the \code{complete\_sa} object]{Plot of the \code{complete\_sa} object. Nearest-neighbor positions have strong interactions (with negative potential) for different-valued pairs, while position $(4,4)$ has a weaker positive potential. This can be interpreted as nearest-neighbors having more weight when they are equal, while pixel with relative position $(4,4)$ have more weight when they are different.}\label{fig:view_model_selection_ex1}
\end{figure}
\end{CodeChunk}

\begin{CodeChunk}

\begin{CodeInput}
R>  # Threshold-based selection
R> thr_value <- 0.1
R> theta_vec <- smr_array(complete_sa$theta, "oneeach")
R> selected <- which(abs(theta_vec) > thr_value)
R> 
R> R1 <- candidates[selected]
R> R1
\end{CodeInput}

\begin{CodeOutput}
3 interacting positions.
  rx     ry
   1      0
   0      1
   4      4
\end{CodeOutput}
\end{CodeChunk}

\begin{CodeChunk}
\begin{figure}[h]

{\centering \includegraphics[width=0.25\linewidth]{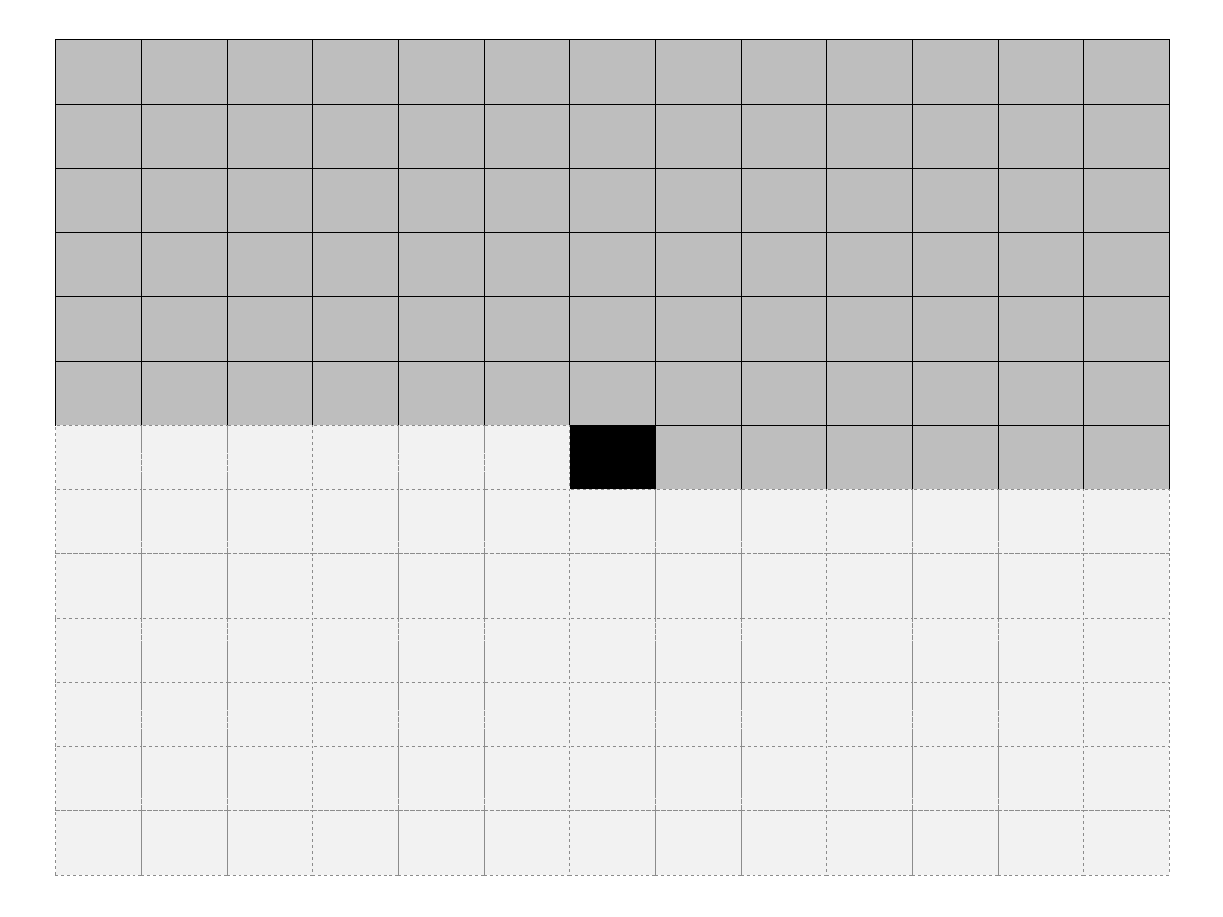} \includegraphics[width=0.25\linewidth]{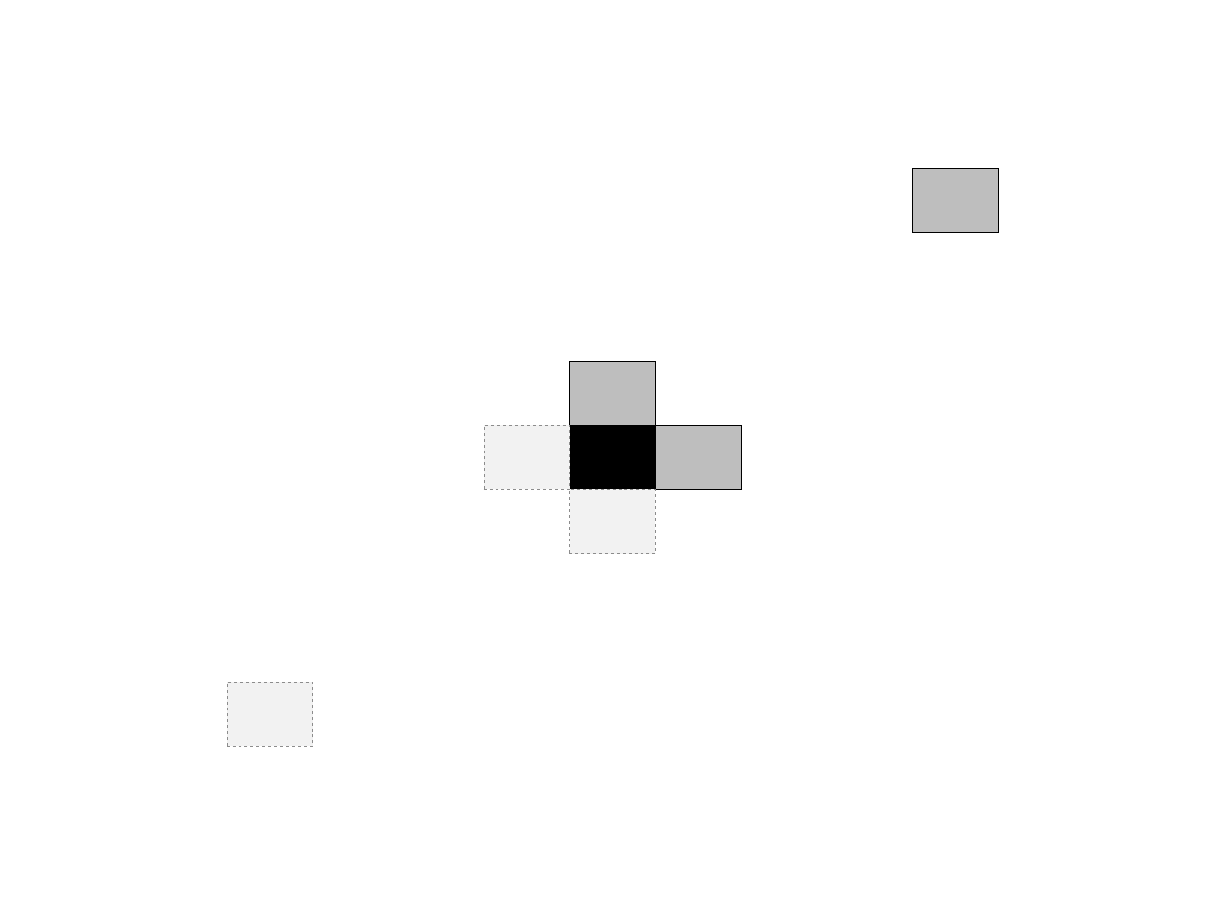} 

}

\caption[Candidate positions for the interaction structure (left) and selected positions (right)]{Candidate positions for the interaction structure (left) and selected positions (right).}\label{fig:show_selection}
\end{figure}
\end{CodeChunk}

\hypertarget{estimating-boldsymboltheta}{%
\paragraph{\texorpdfstring{Estimating
\(\boldsymbol\theta\)}{Estimating \textbackslash boldsymbol\textbackslash theta}}\label{estimating-boldsymboltheta}}

Considering the parameter restriction family \code{"onepar"} and the
selected interaction structure
\(\mathcal{R} = \{(1,0), (0,1), (4,4) \}\), we have a model with 3 free
parameters. A 3-dimensional optimization problem is simple enough to be
solved using the built-in pseudo-likelihood optimization function. We
also fit the model via Stochastic Approximation, now only considering
the selected interaction structure, for comparison. The obtained results
are compared with the \code{summary()} method for the \code{mrfout}
class and presented below.

\begin{CodeChunk}

\begin{CodeInput}
R> pl <- fit_pl(field1, R1, family = "oneeach")
R> summary(pl)
\end{CodeInput}

\begin{CodeOutput}
Model adjusted via Pseudolikelihood 
Image dimension: 150 150 
2 colors, distributed as:
     0      1        
 11083  11417 

Interactions for different-valued pairs: 
Position|  Value  Rel. Contribution
   (1,0)| -0.993  1.000 ***
   (0,1)| -1.021  0.995 ***
   (4,4)|  0.183  0.735 **
\end{CodeOutput}
\end{CodeChunk}

\begin{CodeChunk}

\begin{CodeInput}
R> sa <- fit_sa(field1, R1, family = "oneeach", 
+   gamma_seq = seq(from = 1, to = 0, length.out = 300))
R> summary(sa)
\end{CodeInput}

\begin{CodeOutput}
Model adjusted via Stochastic Approximation 
Image dimension: 150 150 
2 colors, distributed as:
     0      1        
 11083  11417 

Interactions for different-valued pairs: 
Position|  Value  Rel. Contribution
   (1,0)| -0.964  1.000 ***
   (0,1)| -0.983  0.987 ***
   (4,4)|  0.183  0.756 ***
\end{CodeOutput}
\end{CodeChunk}

\hypertarget{evaluating-the-fitted-model}{%
\paragraph{Evaluating the fitted
model}\label{evaluating-the-fitted-model}}

To evaluate how well the estimated parameters fit the data, we generate
a new sample from the fitted model. Figure \ref{fig:show_sampled} shows
the original image and the image simulated from the fitted model for
comparison. The patterns created are visually very similar. Therefore
the MRF model fitted successfully describes the characteristics of the
data and is capable of synthesizing new images with the same texture
pattern.

\begin{CodeChunk}

\begin{CodeInput}
R> z_sim <- rmrf2d(dim(field1), R1, pl$theta)
\end{CodeInput}
\end{CodeChunk}

\begin{CodeChunk}
\begin{figure}[h]

{\centering \includegraphics[width=0.49\linewidth]{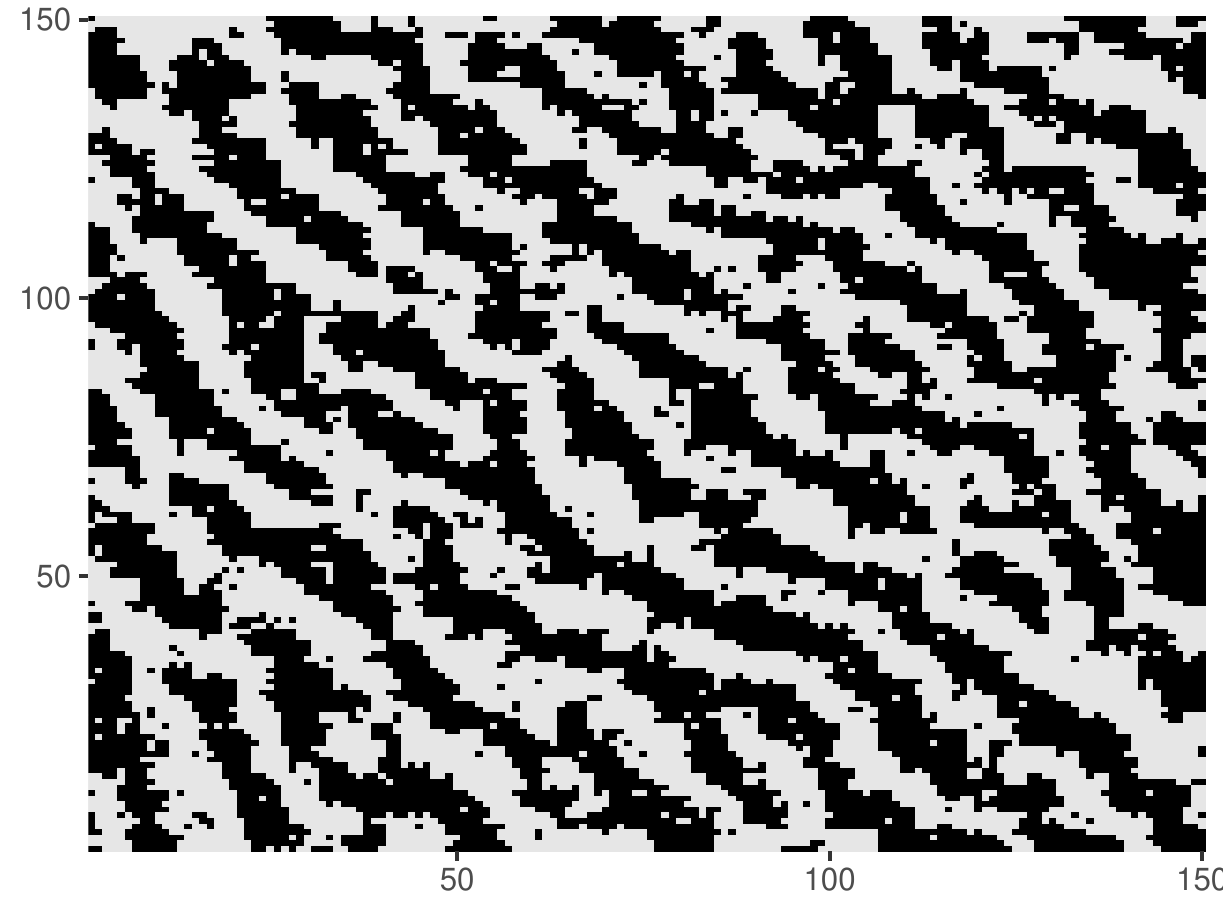} \includegraphics[width=0.49\linewidth]{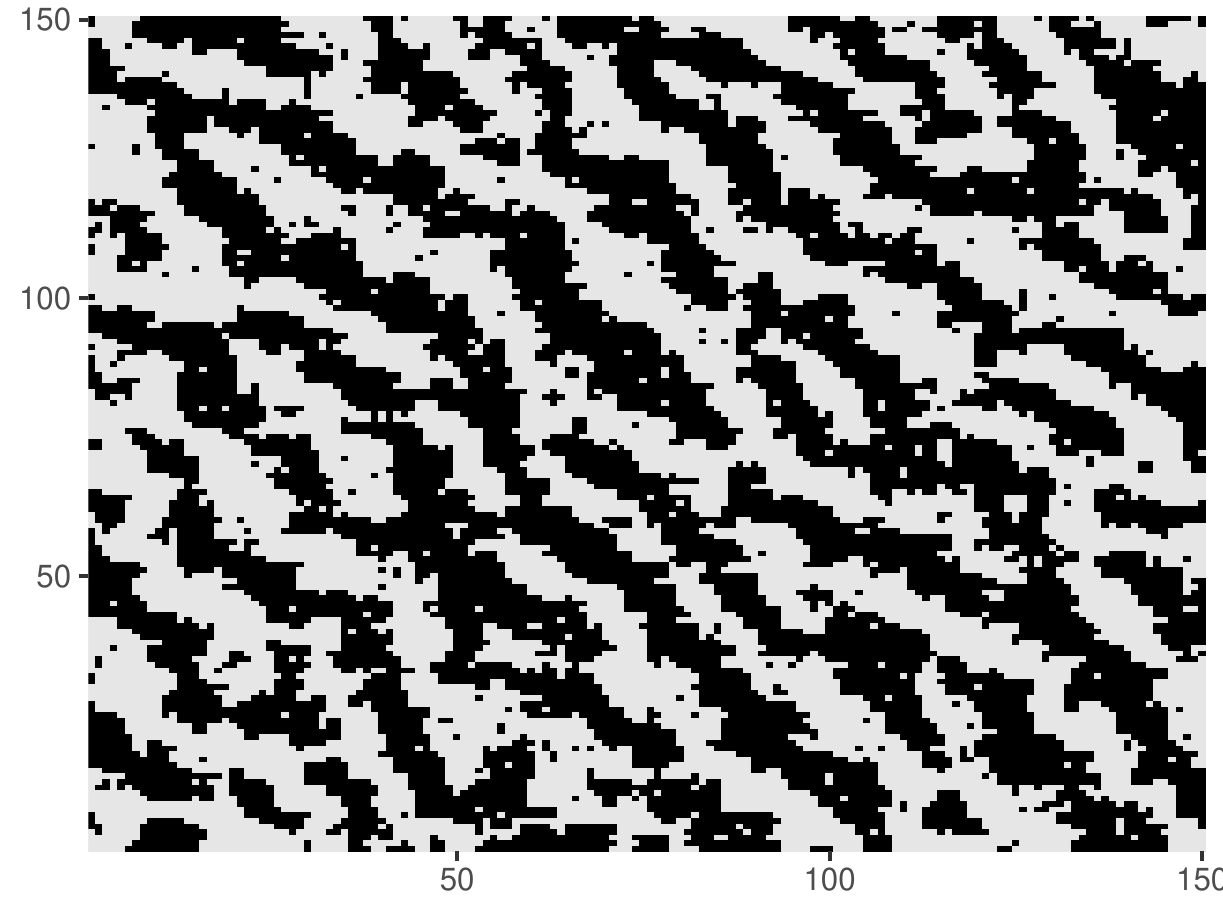} 

}

\caption[Original data (left) and random field simulated from the fitted model (right)]{Original data (left) and random field simulated from the fitted model (right).}\label{fig:show_sampled}
\end{figure}
\end{CodeChunk}

\hypertarget{example-2-image-segmentation-of-a-hidden-mrf-with-spatial-fixed-effect}{%
\subsection{Example 2: Image segmentation of a Hidden MRF with spatial
fixed
effect}\label{example-2-image-segmentation-of-a-hidden-mrf-with-spatial-fixed-effect}}

\hypertarget{description-1}{%
\paragraph{Description}\label{description-1}}

The data available in the object \code{hfield1} in the package will be
used to illustrate the use of \pkg{mrf2d} for Gaussian mixtures driven
by Hidden MRFs. It consists of an image with continuous-valued pixels
ranging from \(0.3\) to \(15.2\). A pattern similar to the previous
example can be observed with the addition of a continuous noise.

\begin{CodeChunk}

\begin{CodeInput}
R> data("hfield1", package = "mrf2d")
R> cplot(hfield1)
\end{CodeInput}
\begin{figure}[ht]

{\centering \includegraphics[width=0.49\linewidth]{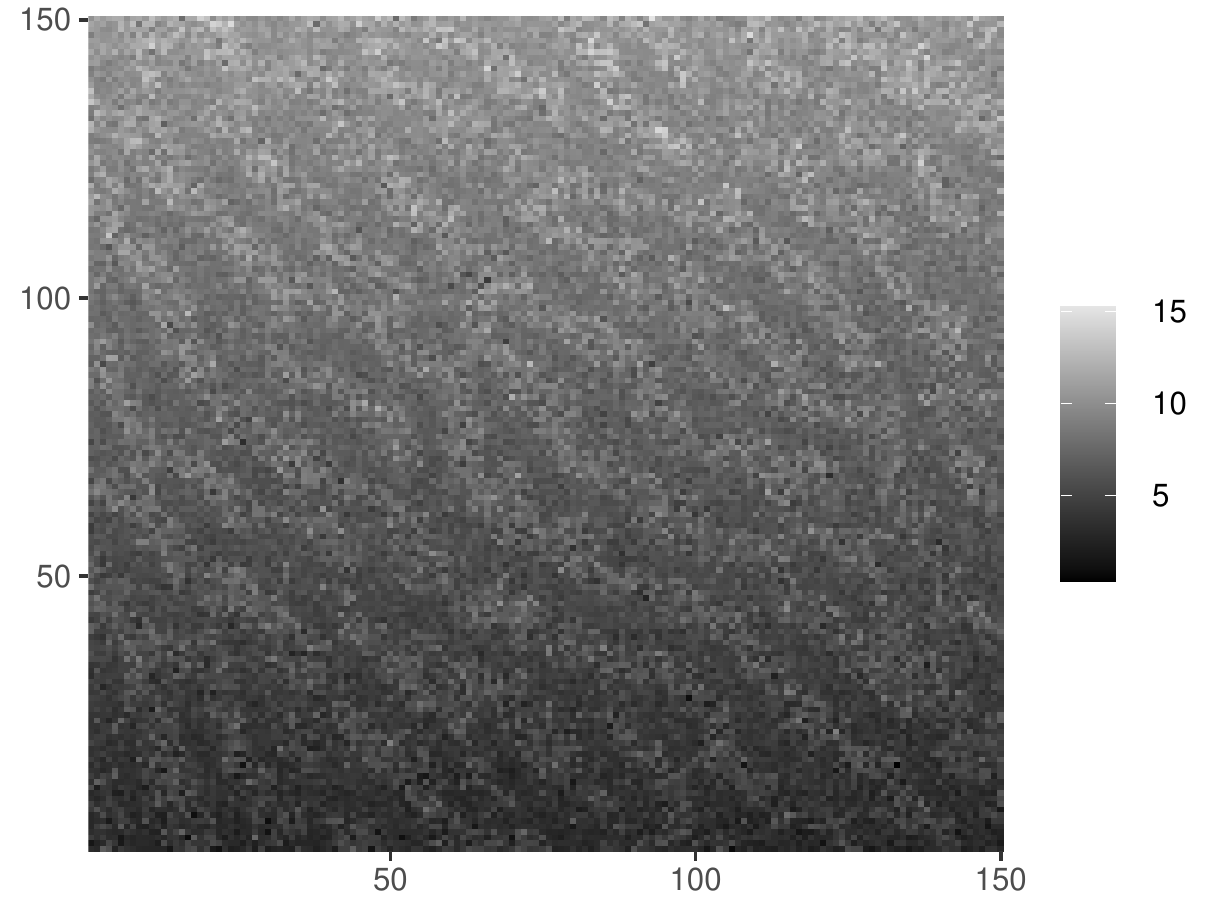} 

}

\caption[Image data for \code{hfield1}]{Image data for \code{hfield1}.}\label{fig:view_hfield1}
\end{figure}
\end{CodeChunk}

We consider the image as a latent Hidden MRF and a random noise which
the distribution on each pixel distribution depends its label. The main
goal in this type of data is to recover the segmentation of the
underlying pixel labels, as an image segmentation problem \citep[for
example]{li2009sar, shah2015automated}. This is the typical problem
where Gaussian mixtures driven by hidden Markov random fields are suited
for.

\hypertarget{fitting-a-hidden-mrf-with-no-fixed-effect}{%
\paragraph{Fitting a Hidden MRF with no fixed
effect}\label{fitting-a-hidden-mrf-with-no-fixed-effect}}

The built-in function for fitting Hidden MRFs (\code{fit_ghm()}), just
like most algorithms used for Gaussian mixtures driven by HMRFs,
considers the distribution of the underlying field as a hyper-parameter
specified a priori. In this example, because a pattern similar to the
one observed in the previous example can be seen forming two cluster of
gray levels locally across the image, we will reuse the model fitted in
Example 1 as the MRF distribution.

We fit a HMRF model to the data using the \code{fit_ghm} function. The
mixture parameters estimates are below and the resulting segmentation is
presented in Figure \ref{fig:hidden_linear_view}(b).

\begin{CodeChunk}

\begin{CodeInput}
R> hmrf_nofixed <- fit_ghm(hfield1, mrfi = R1, theta = pl$theta)
R> summary(hmrf_nofixed)
\end{CodeInput}

\begin{CodeOutput}
Gaussian mixture model driven by Hidden MRF fitted by EM-algorithm.
Image dimensions: 150 150 
Predicted mixture component table:
     0      1        
 11031  11469 
Number of covariates (or basis functions): 0 
Interaction structure considered: (1,0) (0,1) (4,4) 

Mixture parameters:
 Component     mu  sigma 
         0   5.30   1.40 
         1   9.20   1.46 

Model fitted in 5 iterations.
\end{CodeOutput}
\end{CodeChunk}

The labels in the segmentation follow the expected pattern only in the
middle part of the image. Two large clusters without the pattern appear
at the upper and lower parts of the image, what indicates there might be
some missing spatial information not included in the model.

\hypertarget{adding-a-polynomial-trend-as-fixed-effect}{%
\paragraph{Adding a polynomial trend as fixed
effect}\label{adding-a-polynomial-trend-as-fixed-effect}}

A HMRF model without covariates has an intrinsic assumption that the
mean values of pixel intensities given its labels are homogeneous along
the image region. This is not the case for the considered data, as a
vertical gradient effect can be observed.

In order to correct this spatial effect not captured in the model, we
include spatial covariates, in form of polynomial functions of pixel
positions \((i_1, i_2)\) as fixed effect. These covariates can be
specified in \code{fixed_fn} argument of the function and a (centered)
polynomial can be created with the \code{polynomial_2d()} function from
\pkg{mrf2d}.

In this example, we include all terms of a two-dimensional centered
polynomial which maximum degree of the components of its terms are
\(3\), i.e.,

\begin{equation}
p(i_1, i_2) = \sum_{d_1 = 0}^3\sum_{d_2 = 0}^3 \beta_{i_1, i_2} 
\left(i_1 - c_1 \right)^{d_1}
\left(i_2 - c_2 \right)^{d_2},
\end{equation} where the centering position \((c_1, c_2)\) is the middle
pixel position of the image.

\begin{CodeChunk}

\begin{CodeInput}
R> hmrf_poly <- fit_ghm(hfield1, mrfi = R1, theta = pl$theta,
+   fixed_fn = polynomial_2d(c(3,3), dim(hfield1)))
R> summary(hmrf_poly)
\end{CodeInput}

\begin{CodeOutput}
Gaussian mixture model driven by Hidden MRF fitted by EM-algorithm.
Image dimensions: 150 150 
Predicted mixture component table:
     0      1        
 11559  10941 
Number of covariates (or basis functions): 15 
Interaction structure considered: (1,0) (0,1) (4,4) 

Mixture parameters:
 Component     mu  sigma 
         0   6.78   0.62 
         1   7.81   1.19 

Model fitted in 5 iterations.
\end{CodeOutput}
\end{CodeChunk}

The code chunk below illustrates how the resulting fields available in
the function output can be visualized. The results are presented in
Figure \ref{fig:hidden_linear_view}.

\begin{CodeChunk}

\begin{CodeInput}
R> cplot(hfield1)
R> dplot(hmrf_nofixed$Z_pred, legend = TRUE)
R> cplot(hmrf_poly$fixed)
R> dplot(hmrf_poly$Z_pred, legend = TRUE)
\end{CodeInput}
\end{CodeChunk}

\begin{CodeChunk}
\begin{figure}[ht]

{\centering \includegraphics[width=0.49\linewidth]{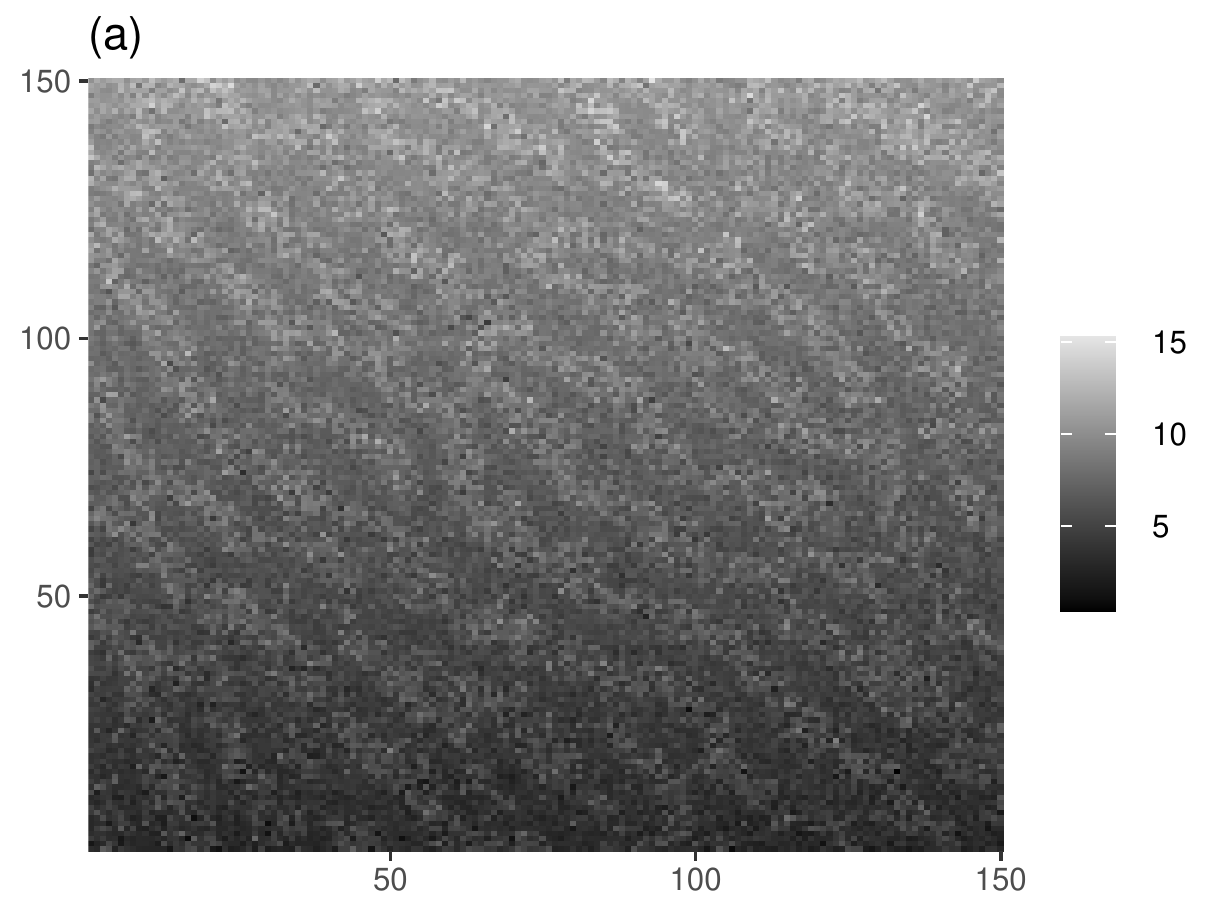} \includegraphics[width=0.49\linewidth]{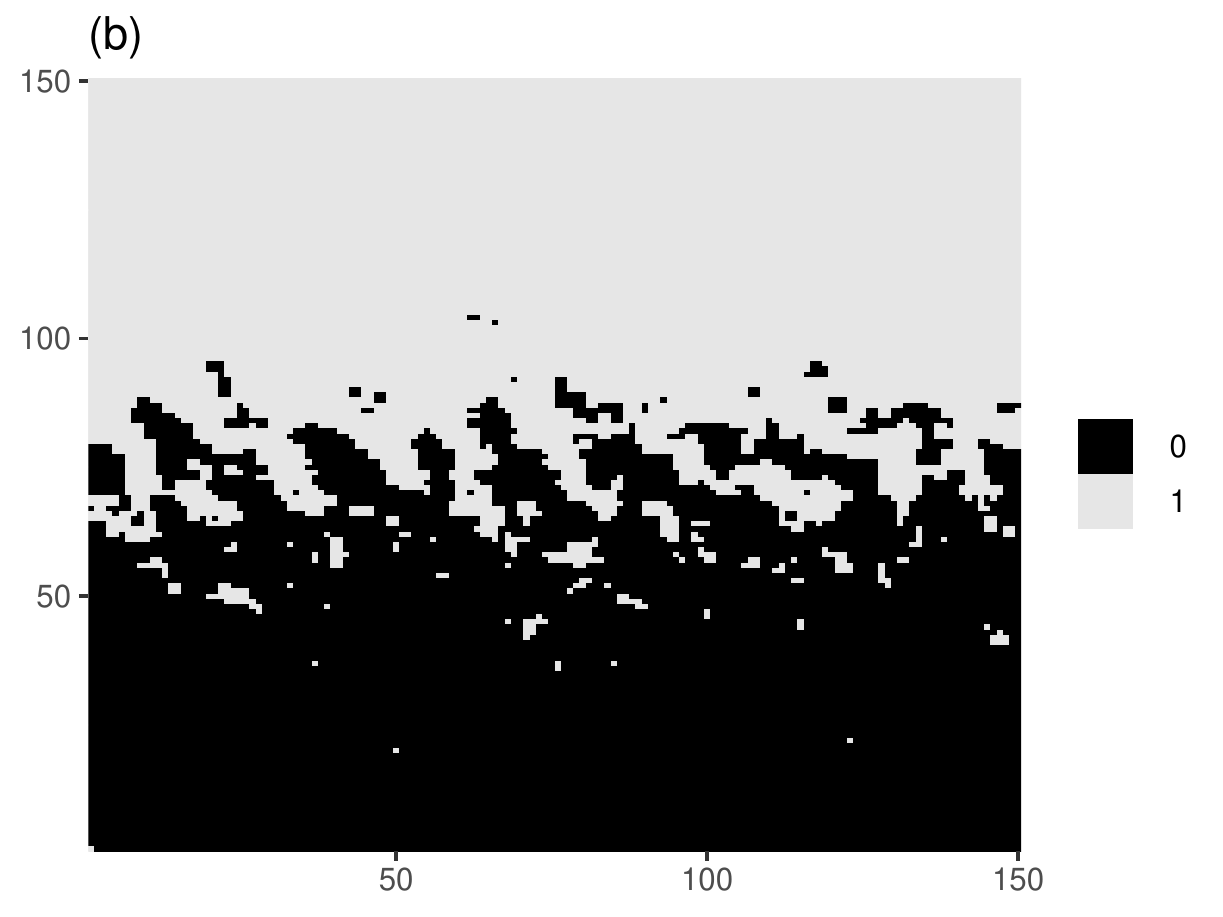} \includegraphics[width=0.49\linewidth]{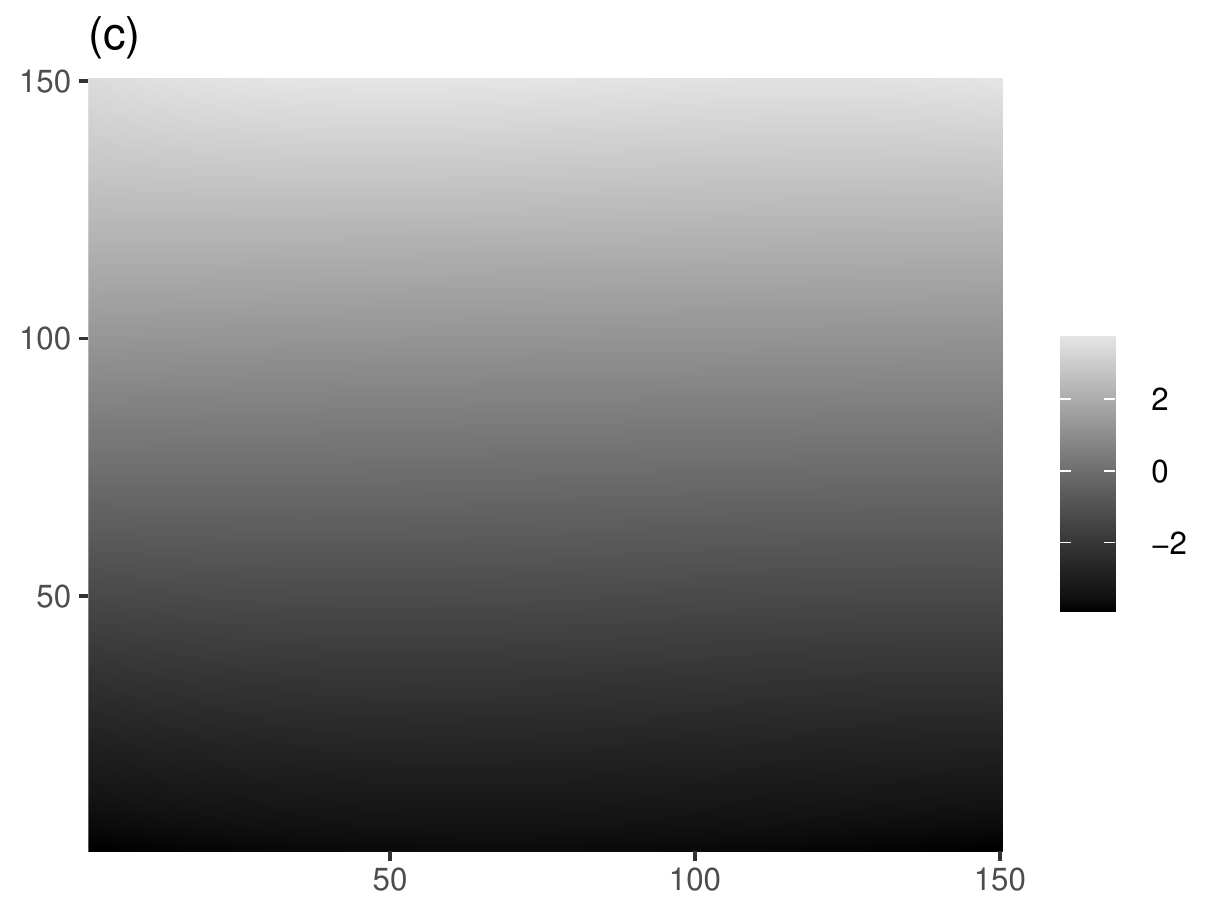} \includegraphics[width=0.49\linewidth]{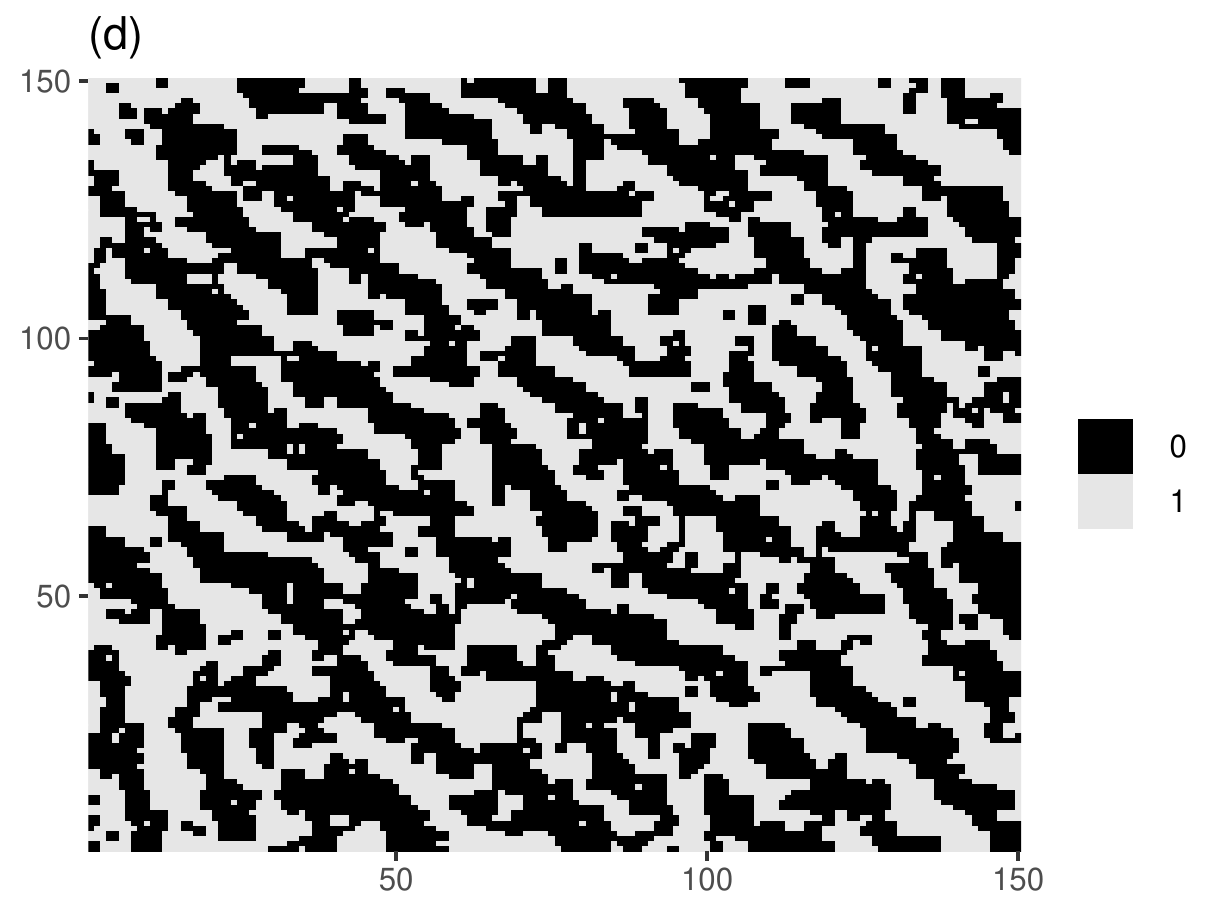} 

}

\caption[Results from the Hidden MRF fits]{Results from the Hidden MRF fits: (a) the original image data, (b) segmentation obtained without adding a polynomial effect, (c) polynomial fitted as a fixed effect, (d) image segmentation when the polyomial effect is included.}\label{fig:hidden_linear_view}
\end{figure}
\end{CodeChunk}

This example highlights the two features of \pkg{mrf2d} that are not
available in other packages: The possibility to specify a distribution
for the underlying field that is more flexible than a simple Potts model
and the option to include covariates (in this example, the polynomial
trend) that are estimated in the simultaneously to the mixture
parameters, allowing us to prevent undesired effects in the segmentation
results.

\hypertarget{example-3-neuroimaging-segmentation-with-bold5000-data}{%
\subsection{Example 3: Neuroimaging segmentation with BOLD5000
data}\label{example-3-neuroimaging-segmentation-with-bold5000-data}}

Neuroimaging is one of the most frequent applications of HMRF models
\citep{zhang2001segmentation, shah2015automated}. We illustrate a brain
magnetic resonance image segmentation using a sample of the BOLD5000
dataset \citep{ref_bold5000} available in the \code{bold5000} object in
the package.

\begin{CodeChunk}

\begin{CodeInput}
R> data("bold5000", package = "mrf2d")
R> cplot(bold5000)
\end{CodeInput}
\begin{figure}[ht]

{\centering \includegraphics[width=0.49\linewidth]{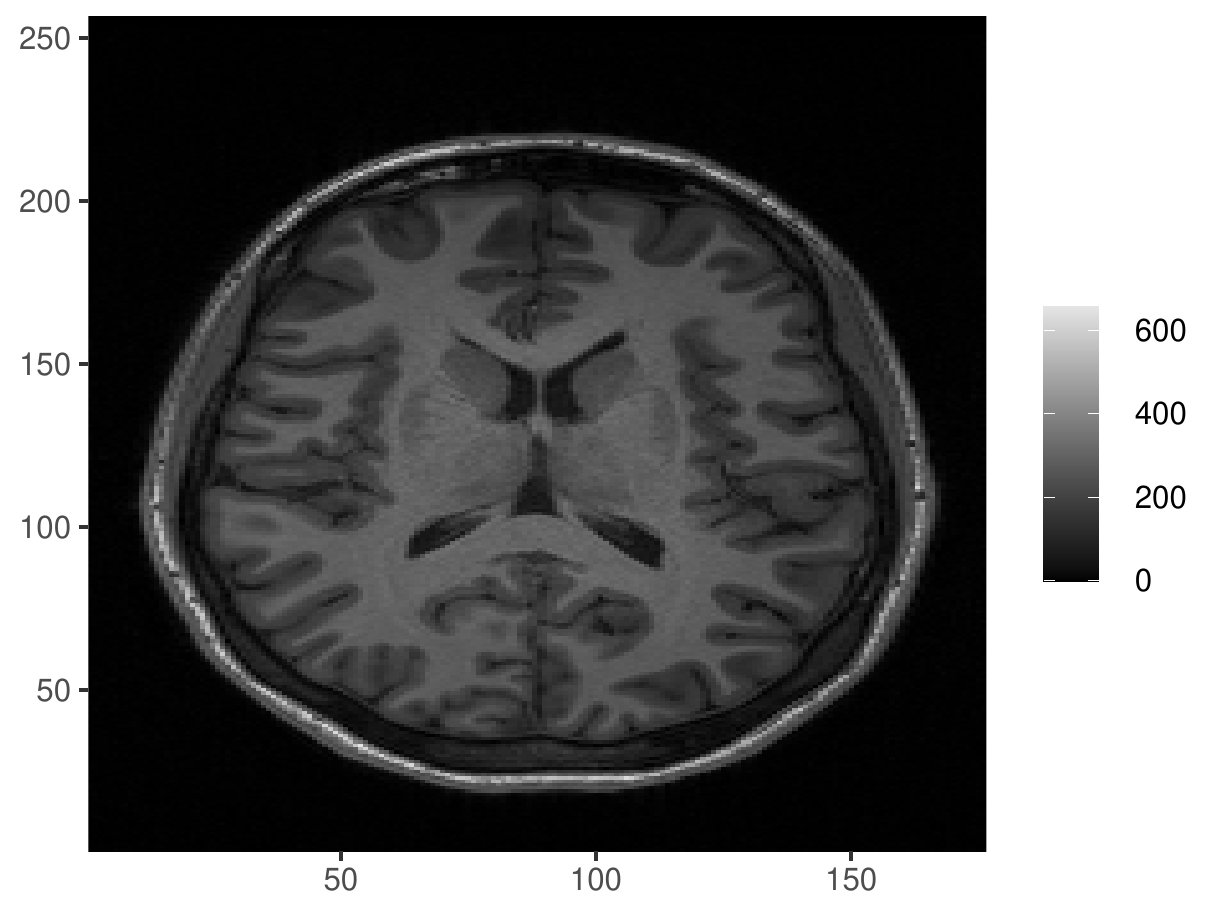} 

}

\caption[Brain magnetic resonance image in \code{bold5000} data]{Brain magnetic resonance image in \code{bold5000} data.}\label{fig:view_brain}
\end{figure}
\end{CodeChunk}

Our main goal in this problem is to segment the brain image into large
regions that corresponds to different elements, like a background,
bones, fat, grey matter, white matter, etc.

The most common approach for the segmentation using HMRFs is to consider
a simple Potts model (nearest-neighbor interaction structure
\(\mathcal{R} = \{ (1,0), (0,1) \}\) and the \code{"onepar"} parameter
restriction family. The potential associated with different-valued pairs
controls, as well as the number of components are considered fixed a
priori and will not be discussed in this paper. For the purpose of
illustration, we use 4 components (\(C = 3\)) and the value \(-1\) for
the potentials of different-valued pairs.

\begin{CodeChunk}

\begin{CodeInput}
R> Rnn <- mrfi(1)
R> theta_nn <- expand_array(-1, family = "onepar", C = 3, mrfi = Rnn)
R> theta_nn
\end{CodeInput}

\begin{CodeOutput}
, , (1,0)

   0  1  2  3
0  0 -1 -1 -1
1 -1  0 -1 -1
2 -1 -1  0 -1
3 -1 -1 -1  0

, , (0,1)

   0  1  2  3
0  0 -1 -1 -1
1 -1  0 -1 -1
2 -1 -1  0 -1
3 -1 -1 -1  0
\end{CodeOutput}
\end{CodeChunk}

We add a constraint that all variance parameters of the mixture
components must be equal by setting the \code{equal_vars} parameter to
\code{TRUE}. This improves the results in this problem by preventing
some of the mixture components to be estimated with too high variance,
what may causes pixels with values too high and too low to be predicted
of the same class with high probability.

We also fit an independent Gaussian mixture (by multiplying all the to
potentials by zero) for a comparison with the HMRF model. Segmentation
results are presented in \ref{fig:view_fit_brain} and the parameter
estimates are shown below.

\begin{CodeChunk}

\begin{CodeInput}
R> fit_brain <- fit_ghm(bold5000, Rnn, theta_nn, equal_vars = TRUE) 
R> fit_brain_ind <- fit_ghm(bold5000, Rnn, theta_nn*0, equal_vars = TRUE)
\end{CodeInput}
\end{CodeChunk}

\begin{CodeChunk}

\begin{CodeInput}
R> summary(fit_brain)
\end{CodeInput}

\begin{CodeOutput}
Gaussian mixture model driven by Hidden MRF fitted by EM-algorithm.
Image dimensions: 176 256 
Predicted mixture component table:
     0      1      2      3        
 22867   6813   7401   7975 
Number of covariates (or basis functions): 0 
Interaction structure considered: (1,0) (0,1) 

Mixture parameters:
 Component     mu  sigma 
         0   7.11  28.77 
         1 128.04  28.77 
         2 207.34  28.77 
         3 294.83  28.77 

Model fitted in 14 iterations.
\end{CodeOutput}

\begin{CodeInput}
R> summary(fit_brain_ind)
\end{CodeInput}

\begin{CodeOutput}
Gaussian mixture model driven by Hidden MRF fitted by EM-algorithm.
Image dimensions: 176 256 
Predicted mixture component table:
     0      1      2      3        
 23013   7020   7066   7957 
Number of covariates (or basis functions): 0 
Interaction structure considered: (1,0) (0,1) 

Mixture parameters:
 Component     mu  sigma 
         0   7.51  30.23 
         1 133.10  30.23 
         2 211.72  30.23 
         3 295.33  30.23 

Model fitted in 4 iterations.
\end{CodeOutput}
\end{CodeChunk}

The resulting parameter estimates are not much different when comparing
the independent mixture model and the HMRF, but the segmentation is
cleaner when using the HMRF model, without sparse different-labeled
pixels inside regions.

\begin{CodeChunk}

\begin{CodeInput}
R> dplot(fit_brain$Z_pred, legend = TRUE)
R> dplot(fit_brain_ind$Z_pred, legend = TRUE)
\end{CodeInput}
\begin{figure}[ht]

{\centering \includegraphics[width=0.49\linewidth]{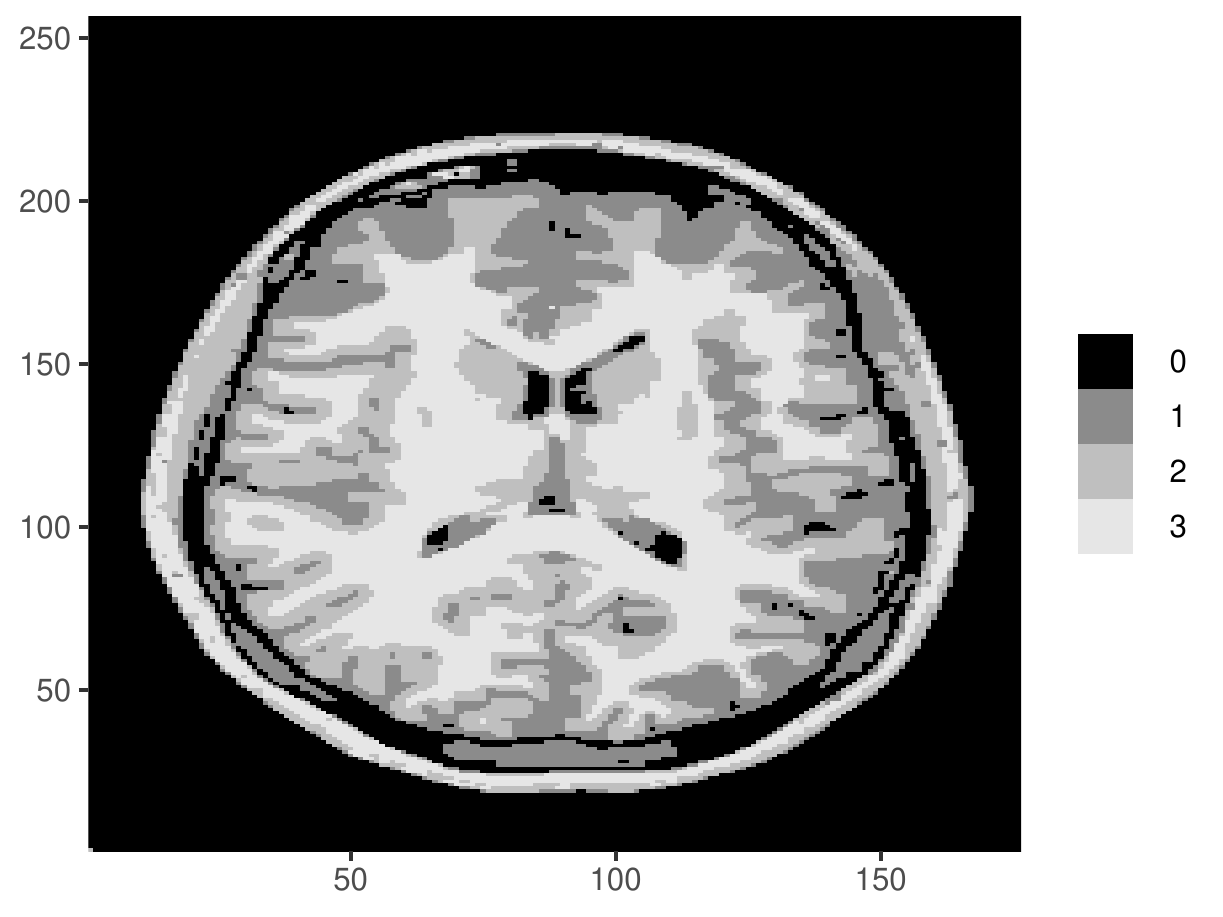} \includegraphics[width=0.49\linewidth]{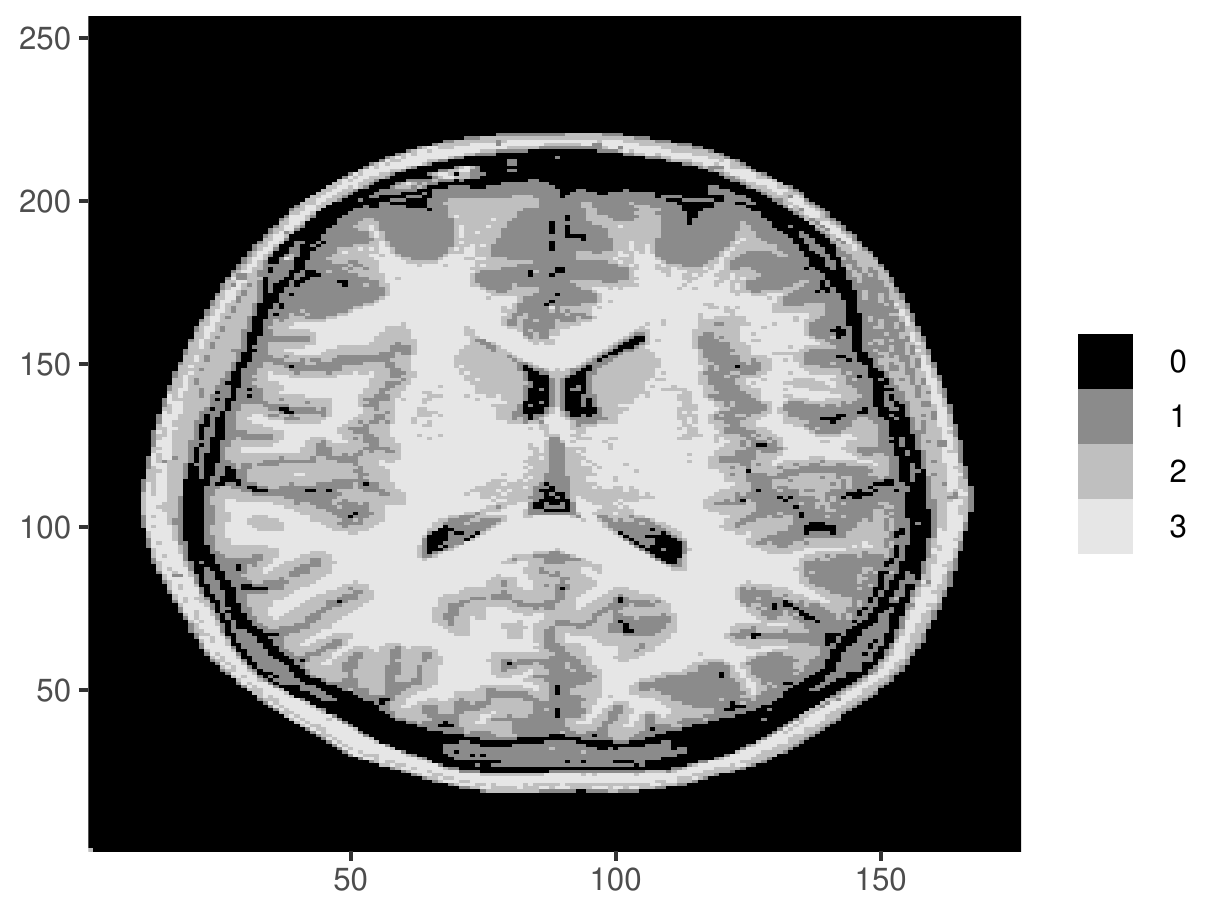} 

}

\caption[Image segmentation predicted by the Hidden MRF fitted (left) and the independent mixture model (right)]{Image segmentation predicted by the Hidden MRF fitted (left) and the independent mixture model (right).}\label{fig:view_fit_brain}
\end{figure}
\end{CodeChunk}

\hypertarget{discussion}{%
\section{Discussion}\label{discussion}}

\pkg{mrf2d} provides a consistent programming interface for statistical
inference in a large class of discrete Markov Random Field models
defined on 2-dimensional lattices. It has an efficient and simple to use
implementation of the main stack of computations used by most of
inference algorithms, as well as complete routines for some commonly
used and more complex estimation methods. The objects used for
representing each model component have been carefully designed and tuned
over several iterations to achieve a balance between performance and
usability in the stable version.

The model featured in the package generalizes Potts model from other
available packages in different ways, such as allowing a flexible
definition of interacting pixel positions and interaction types, with
the drawback that it cannot take advantage of algorithms that require
the setup of a Potts model to improve their efficiency.

We currently have over 160 unit tests supported by the \pkg{testthat}
package \citep{ref_testthat} and more than 90\% of the code covered in
the tests. These tests were designed to verify mathematical correctness
of functions, the behavior of functions with unexpected input and the
consistency of error messages. The package is in constant development
and new tests are added whenever new functionalities are implemented to
ensure its reliability over time.

For these reasons, \pkg{mrf2d} is another important tool for making
statistical inference in images using Markov Random Fields more
accessible, allowing researchers to produce data analysis and implement
new algorithms in \proglang{R} with a simple and consistent framework.

\hypertarget{acknowledgments}{%
\section*{Acknowledgments}\label{acknowledgments}}
\addcontentsline{toc}{section}{Acknowledgments}

This work was funded by Fundação de Amparo à Pesquisa do Estado de São
Paulo - FAPESP grant 2017/25469-2 and by Coordenação de Aperfeiçoamento
de Pessoal de Nível Superior - Brasil (CAPES) - Finance Code 001.

\vfill

\pagebreak

\bibliography{bibliography.bib}
\vfill

\pagebreak
\appendix

\hypertarget{sec:apndx_gg}{%
\section{Customizing visualizations with ggplot2}\label{sec:apndx_gg}}

\hypertarget{random-field-visualization}{%
\subsection{Random Field
Visualization}\label{random-field-visualization}}

The two plotting functions \code{dplot()} and \code{cplot()} return an
object of the \code{ggplot} class, what allows users to produce
customized visualizations by changing scales, legend characteristics,
titles, themes and much more by using the grammar of graphics from the
\code{ggplot2} package \citep{refggplot}.

Random fields (represented by \code{matrix} objects) are transformed
into a \code{data.frame} structure with columns \code{x}, \code{y} and
\code{value}. \code{x} and \code{y} are the indices of the matrix object
while \code{value} maps the pixel-value in that position
(\code{Z[x,y]}).

The plots are constructed using a tile plane with rectangles
(\code{geom_tile()} from \pkg{ggplot2}) with the \code{value} column map
to the \code{fill} aesthetics. In \code{dplot()}, which is used for
finite-valued fields, \code{value} is treated as a factor, while in
\code{cplot()} it is a continuous \code{numeric}. This is the only
difference between the functions and it should be kept in mind when
defining custom color scales.

Figure \ref{fig:dplot_custom} created with the code chunk below shows
examples of customized versions of a random field visualization built
from the same base \code{dplot()} result. Modifications include adding a
title, removing all scale-related information (keeping only the actual
image), using a custom color-scale and changing the legend position,
respectively.

\begin{CodeChunk}

\begin{CodeInput}
R> library("ggplot2")
R> base_plot <- dplot(field1, legend = TRUE)
R> 
R> base_plot + ggtitle("This is a custom title")
R> base_plot + theme_void() + theme(legend.position = "none")
R> base_plot + scale_fill_manual(values = c("red", "blue"))
R> base_plot + theme(legend.position = "bottom")
\end{CodeInput}
\begin{figure}[ht]

{\centering \includegraphics[width=0.45\linewidth]{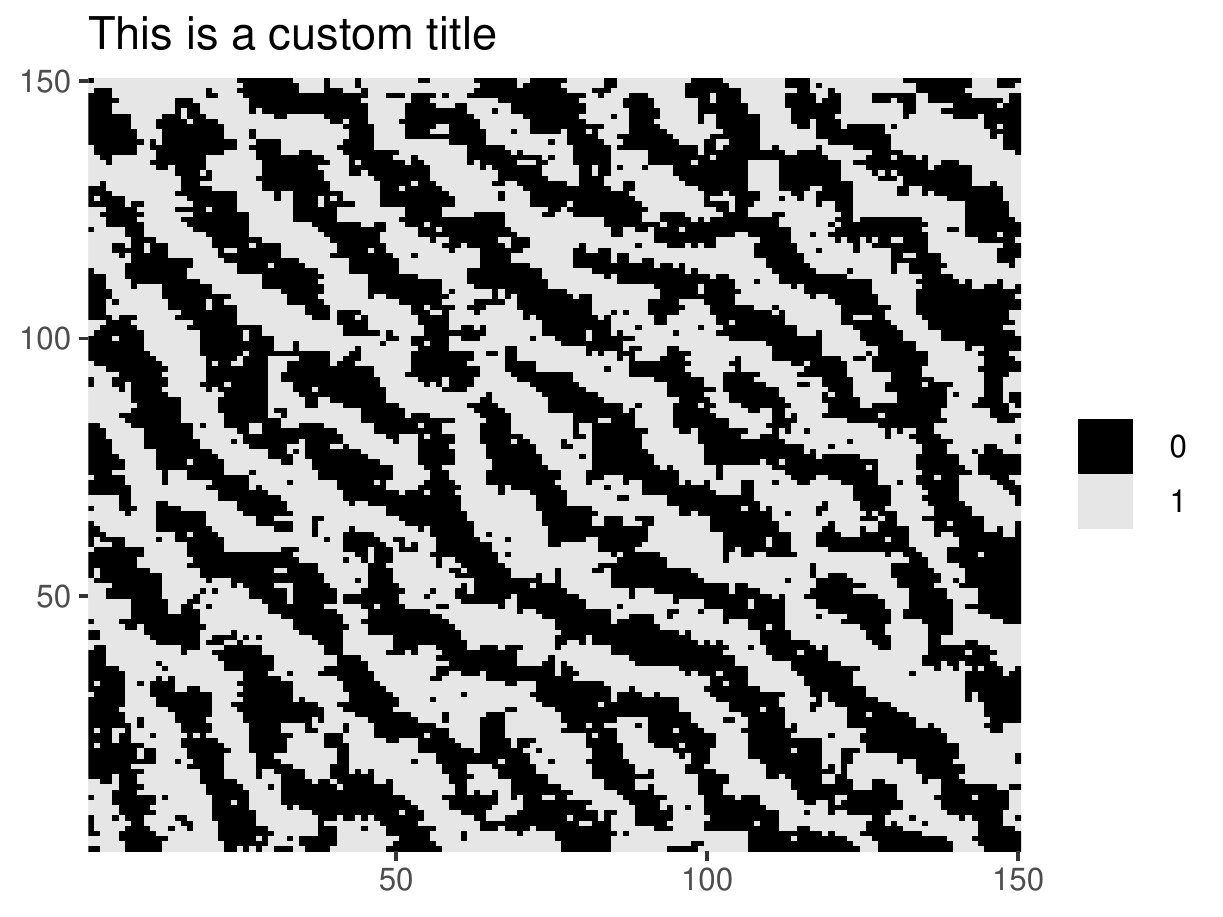} \includegraphics[width=0.45\linewidth]{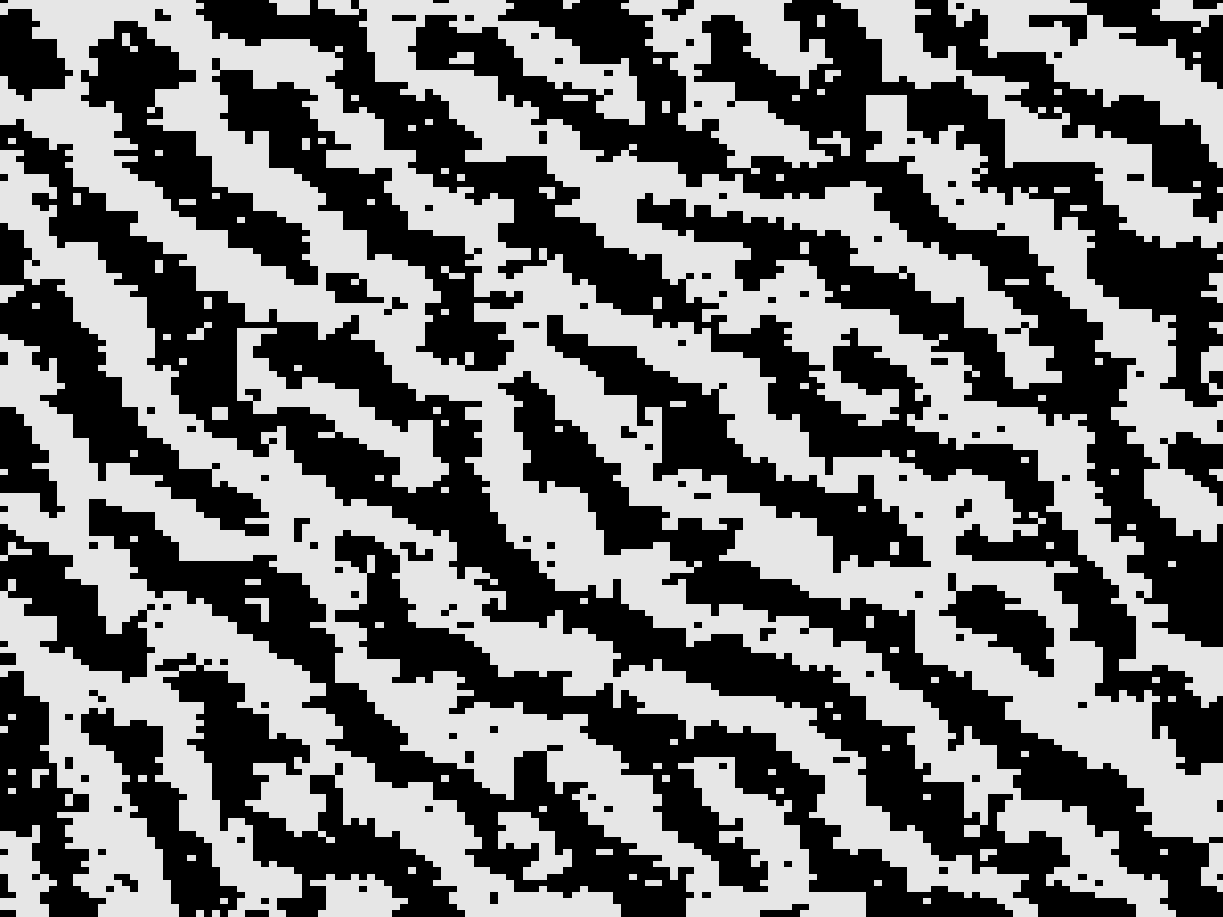} \includegraphics[width=0.45\linewidth]{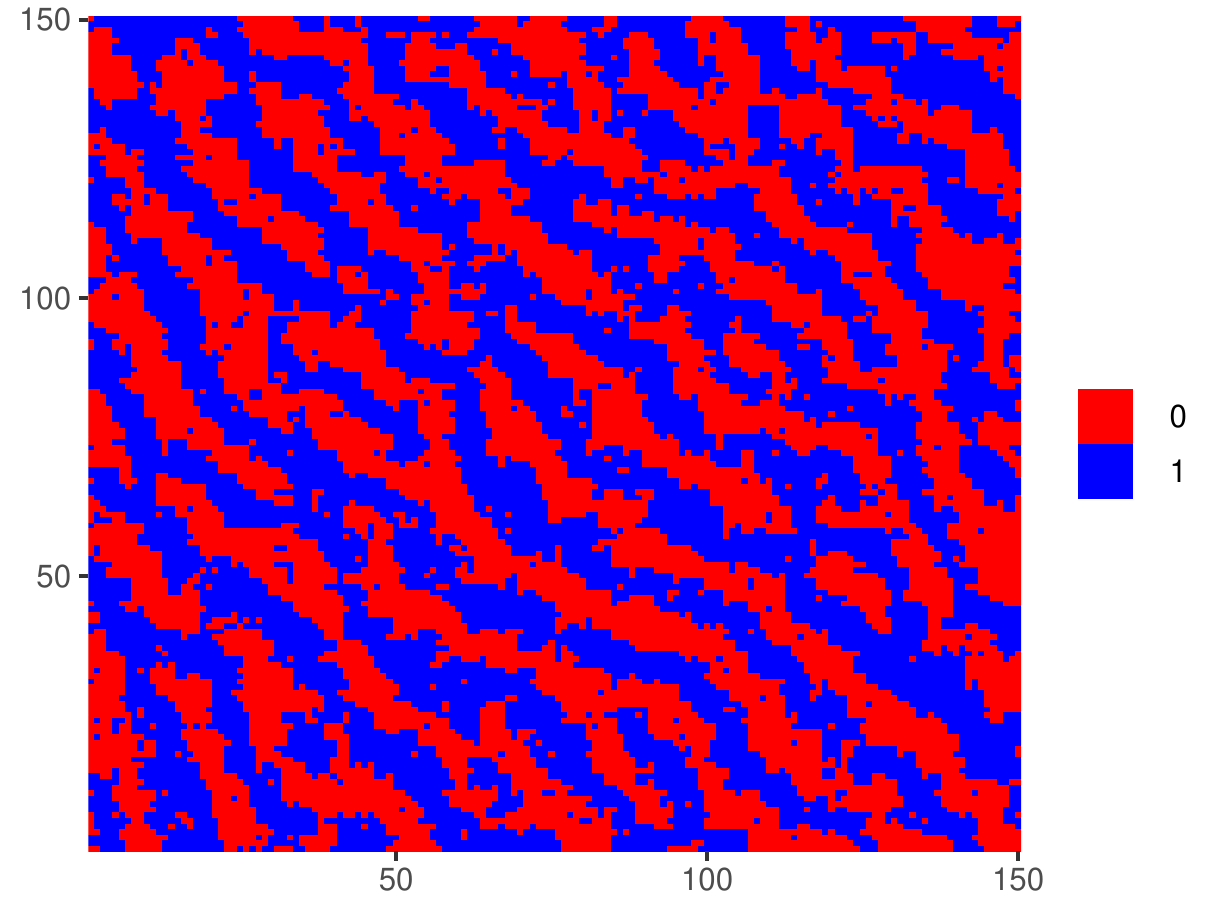} \includegraphics[width=0.45\linewidth]{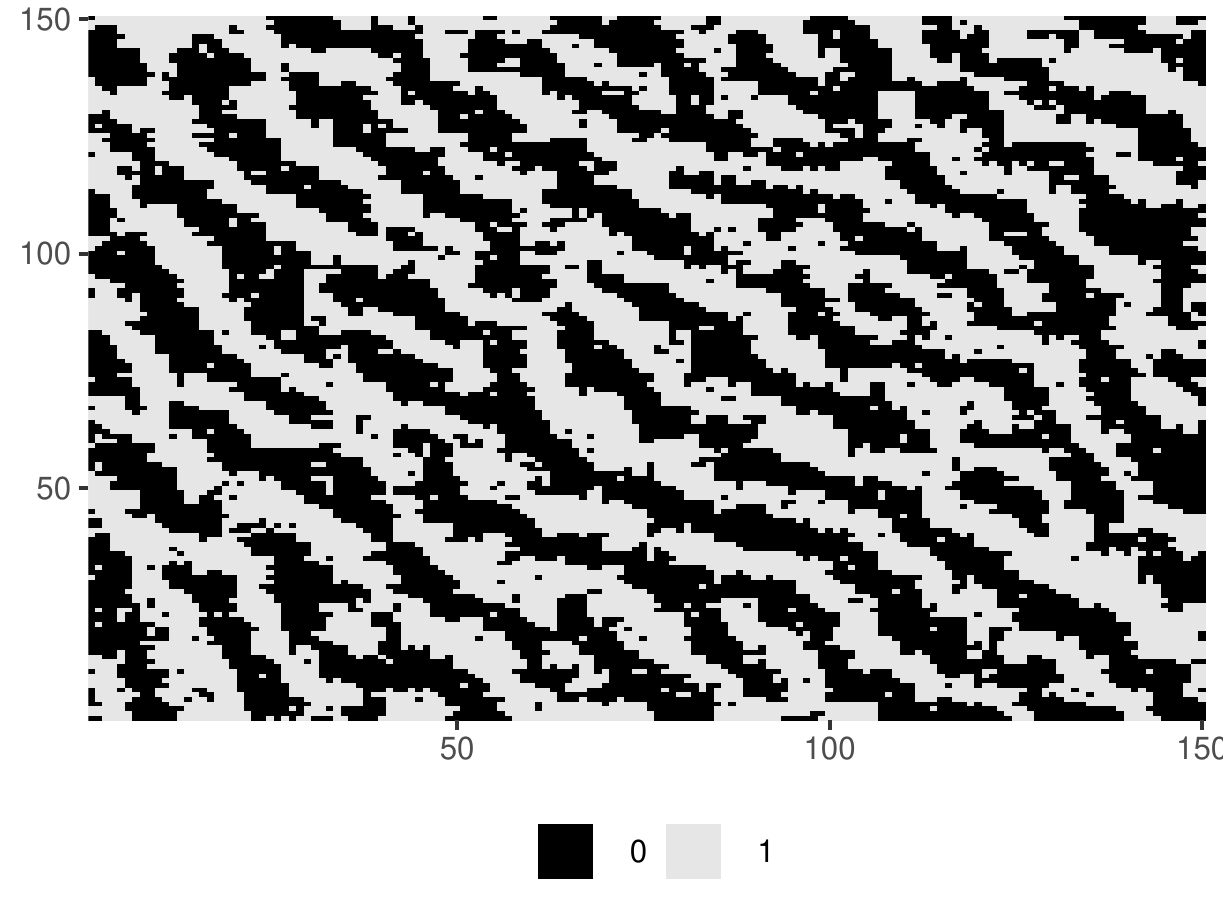} 

}

\caption[Four examples of field visualizations achieved by adding \pkg{ggplot2} layers to a base plot produced in \pkg{mrf2d}]{Four examples of field visualizations achieved by adding \pkg{ggplot2} layers to a base plot produced in \pkg{mrf2d}.}\label{fig:dplot_custom}
\end{figure}
\end{CodeChunk}

\hypertarget{interaction-structure-visualization}{%
\subsection{Interaction Structure
Visualization}\label{interaction-structure-visualization}}

Similarly to the functions used to visualize random fields, the
\code{plot()} method for \code{mrfi} objects. A \code{data.frame} with
columns named \code{rx} and \code{ry} is created internally with the
coordinates of each interacting relative position. These columns are
mapped to the \code{x} and \code{y} axis, respectively and a
\code{geom_tile} is used to produce the plot. The reverse positions are
included automatically with a light-gray color, but this can be
prevented by setting \code{include_opposite = FALSE} in the \code{plot}
call. It also returns a \code{ggplot} object that can be customized.

\begin{CodeChunk}

\begin{CodeInput}
R> mrfi_plot <- plot(mrfi(3) + c(5,1))
R> 
R> # Custom colors
R> mrfi_plot + geom_tile(fill = "orange", color = "blue")
R> # Adding labels with geom_text
R> mrfi_plot + geom_text(aes(label = paste0("(",rx,",",ry,")")))
R> # Custom title
R> mrfi_plot + ggtitle("Add a custom title") + 
+   theme(plot.title = element_text(hjust = 0.5, size = 24))
\end{CodeInput}
\begin{figure}[ht]

{\centering \includegraphics[width=0.32\linewidth]{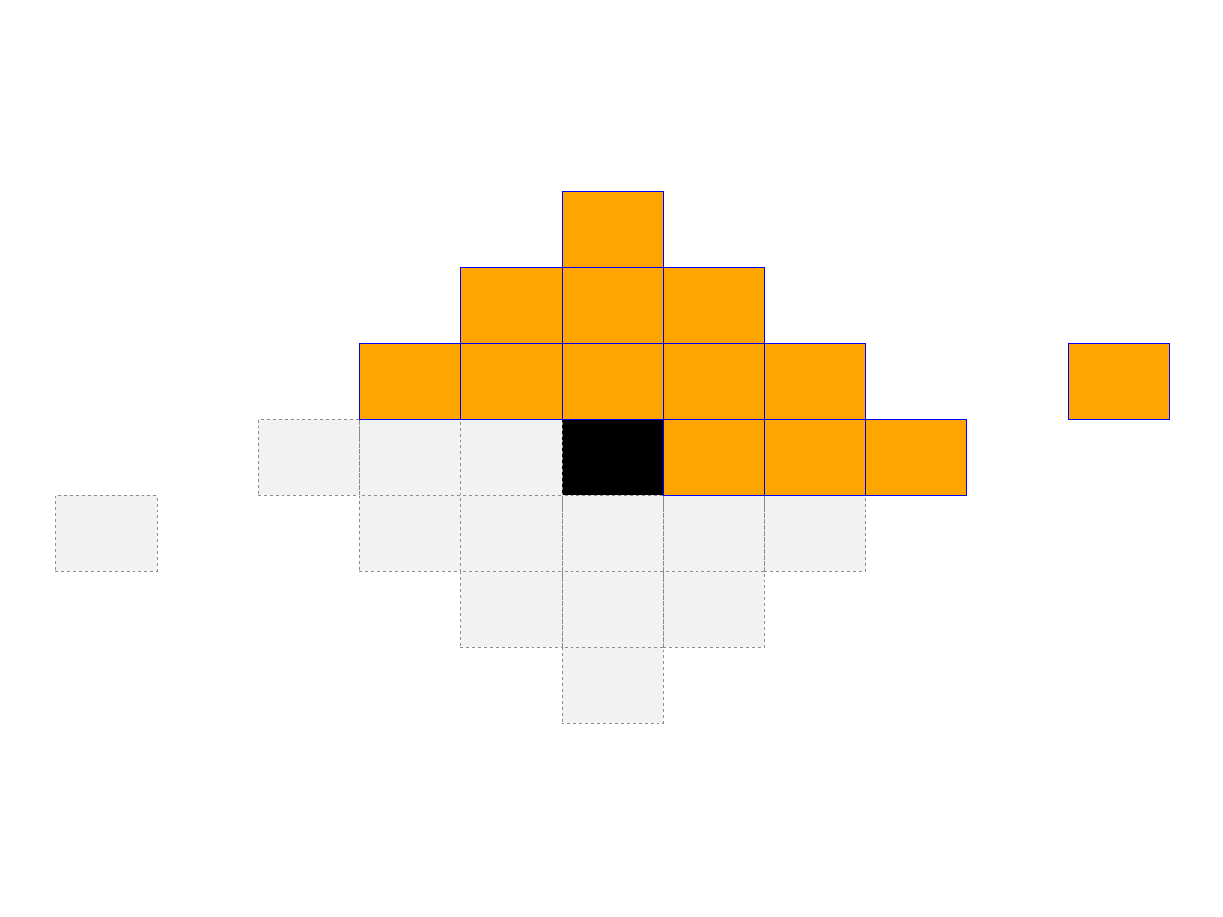} \includegraphics[width=0.32\linewidth]{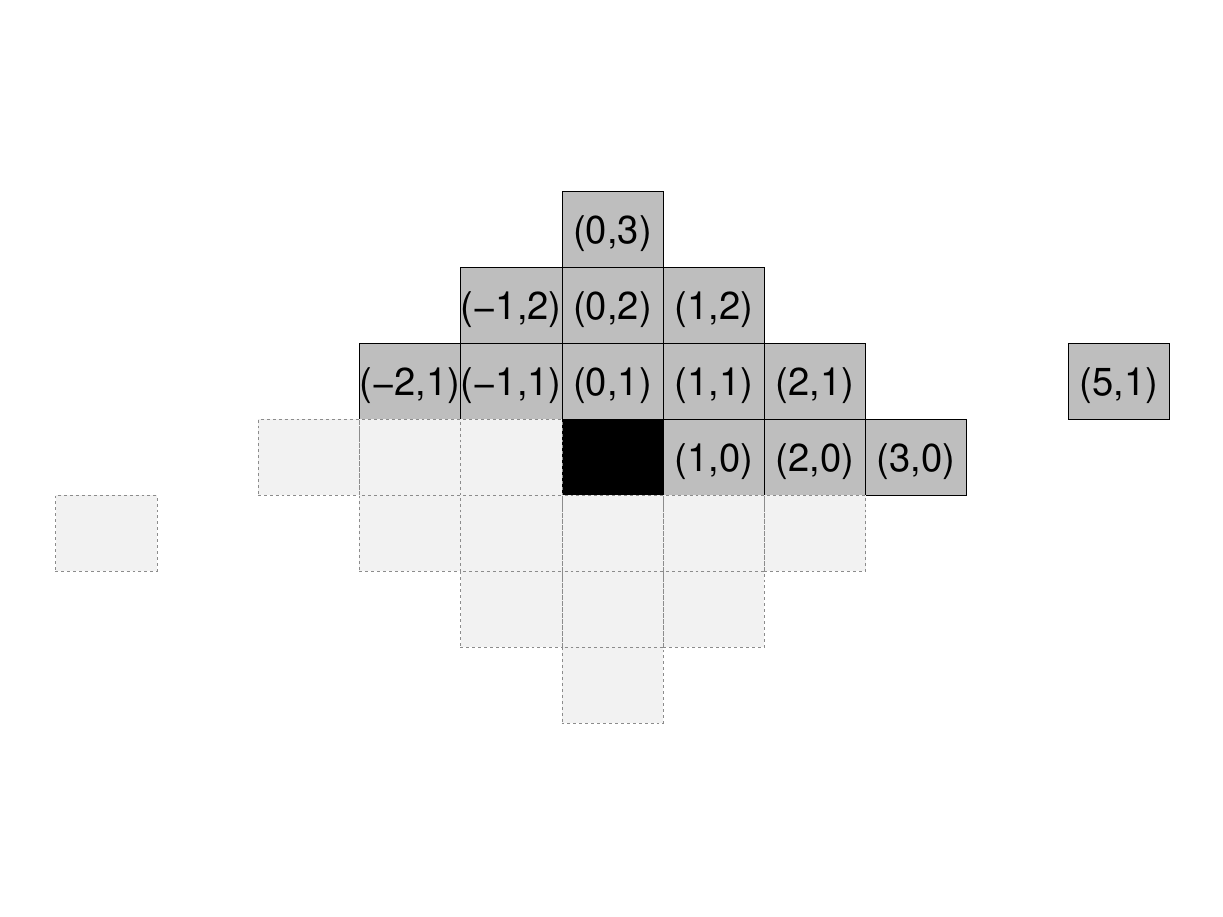} \includegraphics[width=0.32\linewidth]{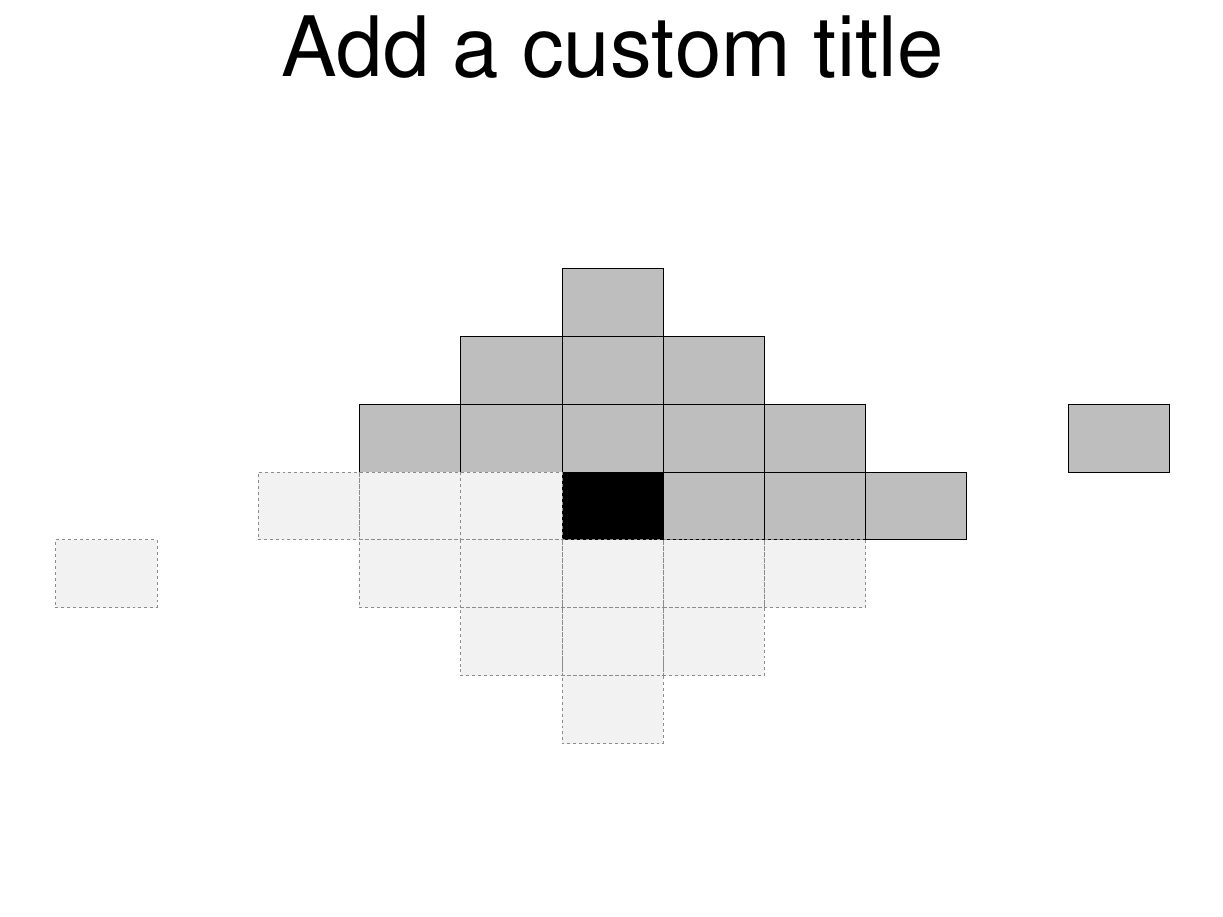} 

}

\caption[Examples of customized visualizations of \code{mrfi} objects]{Examples of customized visualizations of \code{mrfi} objects.}\label{fig:mrfi_custom}
\end{figure}
\end{CodeChunk}

\vfill

\end{document}